\newcommand{\hi}{\ion{H}{i}}
\newcommand{\halpha}{H$\alpha$}
\newcommand{\msun}{{\rm M}_\odot}
\newcommand{\lsunblue}{{\rm L}_{B,\odot}}
\newcommand{\vsun}{V_\odot}
\newcommand{\eso}{\object{ESO\,364-G029}}
\newcommand{\lmc}{\object{LMC}}
\newcommand{\rtf}{R_{25}}

\newcommand{\rhol}{R_{\rm Ho}}
\newcommand{\kms}{km\,s$^{-1}$}
\newcommand{\jykms}{Jy\,km\,s$^{-1}$}
\newcommand{\rotcur}{{\tt ROTCUR}}

\documentclass[]{aa}
\usepackage{txfonts,natbib,graphicx}

\begin{document}
%
%
\title{Optical $BVI$ Imaging and \hi{} Synthesis Observations\\
 of the Dwarf Irregular Galaxy \eso{}}
\titlerunning{Optical and \hi{} Observations of
 \eso{}}
%
%
\author{M.\ B.\ N.\ Kouwenhoven\inst{1,2,3} \and
M.\ Bureau\inst{4} \and
S.\ Kim\inst{5} \and
P.\ T.\ de Zeeuw\inst{2}}
\authorrunning{M.\ B.\ N.\ Kouwenhoven et al.}
%
%
\institute{Department of Physics and Astrophysics, University of Sheffield,
Hicks Building, Hounsfield Road, Sheffield S3 7RH, United Kingdom
(t.kouwenhoven@sheffield.ac.uk)
\and
Sterrewacht Leiden, Leiden University, Niels Bohrweg 2, 2333~CA Leiden, Netherlands
(tim@strw.leidenuniv.nl)
\and
Astronomical Institute Anton Pannekoek, Kruislaan 403, 1098~SJ, Amsterdam,
The Netherlands
\and
Department of Physics, University of Oxford, Denys Wilkinson Building,
Keble Road, Oxford OX1~3RH, United Kingdom
(bureau@astro.ox.ac.uk)
\and
Astronomy \& Space Science Department, Sejong University, 98
Kwangjin-gu, Kunja-dong, Seoul, 143-747, Korea
(sek@sejong.ac.kr)}
\offprints{Thijs Kouwenhoven, \email{t.kouwenhoven@sheffield.ac.uk}}
\date{Received / Accepted}
%
%
\abstract{As part of an effort to enlarge the number of well-studied
Magellanic-type galaxies, we obtained broadband optical imaging and neutral
hydrogen radio synthesis observations of the dwarf irregular galaxy
\eso{}. The optical morphology
characteristically shows a bar-like main body with a one-sided spiral
arm, an approximately exponential light distribution, and offset
photometric and kinematic centers. The \hi{} distribution is mildly
asymmetric
and, although slightly offset from the photometric center, roughly follows
the optical
brightness distribution, extending to over $1.2$ Holmberg radii (where
$\mu_B=26.5$~mag~arcsec$^{-2}$). In particular, the highest \hi{}
column densities closely follow the bar, one-arm spiral, and a third
optical extension. The rotation is solid-body in the inner parts but
flattens outside of the optical extent. The total \hi{} flux
$F_\hi=23.1\pm1.2$~\jykms{}, yielding a total \hi{}
mass $M_\hi=(6.4\pm1.7)\times10^8$~$\msun$ (for a distance
$D=10.8\pm1.4$~Mpc) and a total \hi{} mass-to-blue-luminosity
ratio $M_\hi/L_B=(0.96\pm0.14)$~$\msun / \lsunblue$\ (distance
independent). The \hi{} data suggest a very complex small-scale
\hi{} structure, with evidence of large shells and/or holes,
but deeper observations are required for a detailed
study. Follow-up observations are also desirable for a proper
comparison with the Large Magellanic Cloud, where despite an optical
morphology very similar to \eso{} the \hi{} bears
little resemblance to the optical.
\keywords{Galaxies: individual: \eso{} -- Galaxies:
irregular -- Galaxies: photometry -- Galaxies: kinematics and dynamics
-- Galaxies: structure -- Galaxies: ISM}}
\maketitle
%
%
\section{Introduction\label{sec:intro}}

In the currently popular bottom-up galaxy and structure formation
scenarios, studies of gas-rich low-metallicity dwarf galaxies are
invaluable, as these objects must have dominated the Universe in the
past (e.g.\ \citealt{wf91,kwg93,cole1994}; but see also \citealt{cgo01}).
They
are also expected to be more uniformly distributed than their larger
counterparts (e.g.\ \citealt{ds86}) and may in fact still be the most
common type of object in the Universe (e.g.\
\citealt{mateo1998,ietal99,d99}).

The Magellanic-type spirals (Sm; with rotational symmetry and some
spiral structure) and Magellanic irregulars (Im; asymmetric with no
spiral structure) are at the transition between fully fledged spirals
and true dwarf irregulars (\citealt{v56,v59}). With respect to the latter
two groups of galaxies, Sm and Im galaxies have intermediate
physical properties such as rotation velocity, nuclear concentration,
colour, and neutral hydrogen content. Although the prototype of the
class, the \object{Large Magellanic Cloud} (\lmc{}; classified
as SB(s)m in NED\footnote{NASA/IPAC Extragalactic Database.}), is
amongst the best studied galaxies (e.g.\
\citealt{w97,mahs02,ketal03,bruns2005,marel2006}
and references therein), the general properties of Magellanic systems
are surprisingly poorly known. The classical reference on the subject
remains that of \citet{vf72}. \citet{o91,o94} provides more modern
studies and argues that it is a misconception to consider the
\lmc{} and other galaxies like it as irregular. Magellanic-type
galaxies are characterized by an asymmetric spiral arm connected at
one end to a high surface brightness bar. The bar center is often
offset from the center of the galaxy as defined by the outer optical
isophotes, and the arm normally has a clumpy appearance, presumably
due to triggered star formation. The arm can sometimes be followed
completely around the galaxy and, in some cases, small ``embryonic''
arms are present at the ends of the bar (\citealt{v55}).

Various simulations have shown a that one-armed morphology can develop from
a
strong tidal encounter with a companion (e.g.\
\citealt{bsv86,hb90}). The models generally predict only a short-lived
one-armed spiral structure, implying that a companion galaxy should
generally be present close to any Magellanic galaxy. Using a sample
from the Third Reference Catalog of Bright Galaxies
(\citealt*{vetal91}, hereafter \citeauthor{vetal91}), \citet{o94}
demonstrated that close companions are indeed observed in almost every
case. As for the \lmc{} itself, its interaction with both the
Milky Way and the \object{Small Magellanic Cloud} (\object{SMC}) is
most likely at the origin of its one-arm structure and large-scale
disturbed morphology (e.g.\ \citealt{puetal98,psfgb03} and references
therein).

%
\begin{table}[tb]
 \caption{Basic properties of the dwarf irregular galaxy \eso{}. The left
and middle columns list the different quantities and their values; the right
column lists corresponding references -- (1) NED; (2) HyperLEDA; (3) This
paper; (4) \cite{kkhm04}. }
 \label{tab:phys_par}
 \centering
 \begin{tabular}{p{39mm}p{36mm}p{1mm}}
   \hline
   \hline
   Quantity                                    & Value
& R.\\
   \hline
   Morphological type                          & IB(s)m
& 1\\
   Right ascension (J2000)                     & $\alpha=06^{\rm h}05^{\rm
m}45\fs2$   & 1\\
   Declination (J2000)                         &
$\delta=-33\degr04\arcmin51\arcsec$   & 1\\
   Galactic longitude                          & $l=239\fdg47$
& 1\\
   Galactic latitude                           & $b=-23\fdg36$
& 1\\
   \hline
   Heliocentric radial velocity                & $\vsun=786 \pm 11$~\kms{}
& 2\\
   \quad (from \hi{} measurements)             & $\vsun=784 \pm 2$~\kms{}
& 3\\
   \hline
   Scale length ($B$)                          & $h_B = 50\arcsec \pm
5\arcsec$       & 3\\
   Radius\,($25$\,$B$\,mag\,${\rm arcsec}^{-2}$)   & $\rtf=1\farcm29 \pm
0\farcm09$        & 2\\
                                               & $\rtf=1\farcm27 \pm
0\farcm03$        & 3\\
   Radius\,($26.5$\,$B$\,mag\,${\rm arcsec}^{-2}$) & $\rhol=2\farcm30 \pm
0\farcm08$       & 3\\
   \hline
   Inclination                                 & $i=70\fdg5$
& 2\\
   Axial\,ratio\,($25$\,$B$\,mag\,${\rm arcsec}^{-2}$) & $q_{25} = 0.51 \pm 0.06$
& 2\\
                                               & $q_{25} = 0.54 \pm 0.04$
& 3\\
   Asymmetry                                   & $Q_{\rm asym} = 1.7\%$
& 3\\
   \mbox{Position\,angle\,($25$\,$B$\,mag\,${\rm arcsec}^{-2}$)} & PA $= 55.4\degr$
& 2\\
                                               & PA $= 62\degr \pm 4\degr$
& 3\\
   \hline
   Central surface brightness                  & $\mu_B(0) = 23.3$~mag~${\rm
arcsec}^{-2}$ & 3\\
   Apparent total $B$ mag                      & $B_{\rm T}=13.81 \pm
0.22$~mag        & 2\\
                                               & $B_{\rm T}=13.8 \pm
0.1$~mag         & 3\\
   Corrected apparent $B$ mag                  & $B_{\rm c}=13.8 \pm
0.22$~mag                  & 2\\
                                               & $B_{\rm c}=13.6 \pm
0.1$~mag              & 3\\
   Corrected absolute $B$ mag                  & $M_B=-16.44$~mag
& 2\\
                                               & $M_B=-16.6 \pm 0.3 $~mag
& 3\\
   \hline
   Distance                                    & $D=8.02$~Mpc
& 2\\
                                               & $D=10.8\pm 1.4$~Mpc
& 3\\
						& $D=7.7$~Mpc
& 4\\
   Scale                                       & $1\arcmin=2.3$~kpc
& 2\\
                                               & $1\arcmin=3.1\pm0.4$~kpc
& 3\\
						& $1\arcmin=2.2$~kpc
& 4\\
   \hline
   Total \hi{} mass                            & $M_{\rm
HI}=(6.4\pm1.7)\times10^8$~$\msun$    & 3\\
   \mbox{\hi{}\,mass-to-blue-luminosity\,ratio}         &
\mbox{$M_\hi/L_B=0.96\pm0.14\msun/\lsunblue$} & 3\\
   \hline
   \hline
 \end{tabular}\\
\end{table}

The kinematics of the warp in the \hi{} distribution of the Magellanic-type
galaxy \object{NGC~3109} and the nearby Antlia dwarf galaxy also suggests that the
galaxies had a mild interaction about 1~Gyr ago (\citealt{barnes2001}).
Based on the morphology and kinematics, \cite{bush2004} draw a similar
conclusion for the neighbours \object{NGC~4618} and \object{NGC~4625}.
However, not all Magellanic-type galaxies have an obvious neighbour for a
recent interaction. \cite{bekki2006} suggest that the morphologies of these
apparently isolated galaxies may be explained by a recent interaction with
an (optically) invisible companion. The \hi{} survey performed by
\cite{doyle2005}, however, shows that evidence for the existence of these
``dark galaxies'' is marginal.

In order to address the asymmetry issue,
\cite{wilcots2004} performed an \hi{} survey of a sample of 13~Magellanic
spiral galaxies with apparent optical companions. In their study, they find
that only four of these have confirmed \hi{}-detected neighbours. Their
study also indicates
that the presence of companions near \object{NGC~2537} and \object{UGC~5391} has no effect on
the morphology of these galaxies, and that Magellanic spirals are no more
asymmetric than a random sample in the field. The latter conclusion is
supported by \hi{} observations of the interacting Magellanic spirals
\object{NGC~4618} and \object{NGC~4625} (\citealt{bush2004}).
On the other hand, in their \hi{} study of (possibly) interacting Magellanic
galaxies in the
\object{M81} group, \cite{bureau2004} find no large-scale tidal feature
and no intergalactic \hi{} cloud near the Magellanic dwarf \object{M81dwA}, and
remark
that this galaxy could be a tidal dwarf galaxy (rather than a perturbed
``regular'' dwarf),
resulting from the debris of a past encounter with \object{Holmberg~II}.

Magellanic dwarf galaxies are also known to contain a large number
of shells and holes in their interstellar medium. These structures are
created by winds and supernova explosions of the most massive stars
in star forming regions. Unlike in spiral systems, these structures
are long-lived in dwarf galaxies,
due to solid-body rotation and a lack density of waves.
Well-known dwarfs exhibiting shells and holes include the \lmc{}
(SB(s)m; \citealt{ketal98}), \object{IC~2574} (SAB(s)m;
\citealt{wb99}) and \object{Holmberg~II} (Im; \citealt{pwbr92}; but
see also \citealt{rswr99,bc02}).

A significant amount of neutral hydrogen is known to be present beyond the
optical extent of Magellanic-type galaxies, often up to 2--3 Holmberg radii
(e.g. \object{NGC~925}; \citealt{pisano1998,pisano2000}; \object{DDO~43};
\citealt{simpson2005}), occasionally reaching up to 6~Holmberg radii
(\object{NGC~4449}; \citealt{hunter1998}).
The \hi{} distribution of several Magellanic dwarfs is known to exhibit
further irregularities, such as \hi{} loops surrounding the galaxy
(\object{NGC~4618}; \citealt{bush2004}) and external spurs or blobs (e.g. \object{NGC~5169};
\citealt{muhle2005}; \object{UGCA~98}; \citealt{stil2005}). An S-shaped distortion in
the \hi{} velocity field, possibly indicating counterrotation, has been
observed in \object{NGC~4449} (\citealt{hunter1998}) and \object{DDO~43}
(\citealt{simpson2005}).

The asymmetries and irregularities in the \hi{} distribution of
Magellanic-type galaxies are often associated with a high star formation
rate, such as for \object{IC~10} (e.g. \citealt{wilcots1998}; see also
\citealt{thurow2005}).
Using \halpha{} observations, \cite{wilcots2001} find that, in several
regions of the Magellanic spiral \object{NGC~4214}, the velocity of the ionized gas
is $50-100$~\kms{} higher than that of the \hi{}, indicating massive star
formation. The study of \object{NGC~4449} by \cite{hunter2000} compares \halpha{},
H$_2$, \hi{} and near-infrared emission, and suggests that the different
regions of this galaxies are in different stages of star formation.

\begin{figure}[tb]
 \resizebox{0.9\hsize}{!}{\includegraphics{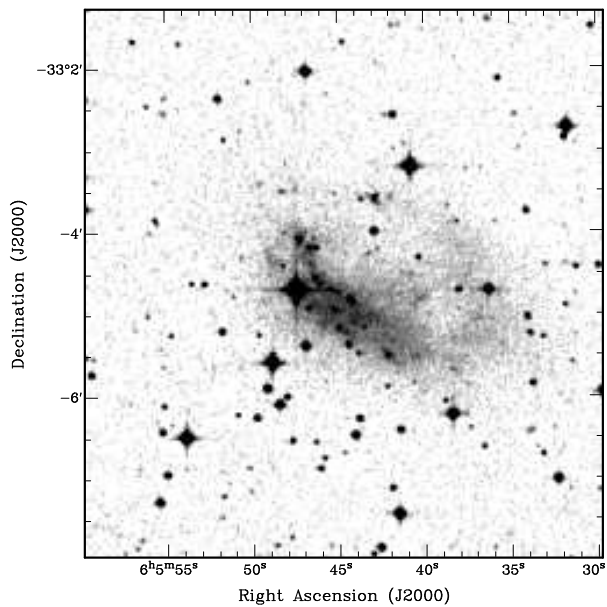}}
 \caption{Optical image of \eso{} from the Digitized Sky Survey (DSS); cf.
our $BVI$~images in Figure~\ref{fig:opt_images}. }
 \label{fig:dss}
\end{figure}

As part of an effort to increase the number of Magellanic dwarfs with
detailed observations, we focus in this paper on the relatively nearby
Magellanic dwarf irregular galaxy \eso{}
(\object{PGC~018396}).
\eso{} was first reported by \cite{holmberg1978}, who classified it as a
``dwarf irregular or emission nebulae'' using the ESO~(B) Atlas.
By combining optical data from the UK~Schmidt plates with Parkes \hi{}
observations, \cite{lhgmw82} concluded that \eso{} is a dwarf irregular
galaxy, although \cite{kk00} designate the object as ``dwarf irregular or
reflection nebula'' in their optical search for nearby dwarf galaxy
candidates in the Southern hemisphere.
\eso{} was identified (and rejected) as a possible companion of
the nearby dwarf galaxy \object{NGC~2188} by \cite{ddd96}.
\cite{zimmermann2001} reported a possible X-ray counterpart to \eso{} using
ROSAT data.
\eso{} appears in a handful of other surveys, such as the Nearby Optical
Galaxies Catalogue (\citealt{gmcp00}), the Catalog of Nearby Galaxies
(\citealt{kkhm04}), the HIPASS\footnote{\hi{} Parkes
All-Sky Survey:\\
http://www.atnf.csiro.au/research/multibeam/release/~.} catalogue
(\citealt{hkk01,metal04,ketal04}), and the surveys for optical/\hi{}
counterparts of \cite{paturel2005} and \cite{doyle2005}.
Although the general properties (such as luminosity and \hi{} content) of
\eso{} are reasonably well constrained by the above-mentioned surveys, the
detailed morphology and kinematics of the galaxy were not studied in detail
previously.
General properties of \eso{} are listed in Table~\ref{tab:phys_par} and an
optical image from the Digitized Sky Survey (DSS) is shown in
Figure~\ref{fig:dss}.

This paper is organized as follows.
Optical observations and surface photometry of \eso{}
are presented in \S~\ref{sec:optical} while \hi{} spectral
line-imaging is discussed in \S~\ref{sec:hi}. In \S~\ref{sec:dist_fluxes}
we constrain the distance to \eso{}. The
discussion in \S~\ref{sec:discussion} focuses on a comparison with the
\lmc{}, the importance of interactions, and the possible presence
of \hi{} shells and holes in \eso{}. Finally, our main conclusions
are briefly summarized in \S~\ref{sec:summary}.
%
%
%
\begin{figure*}[tb]
 \begin{tabular}{ccc}
   \includegraphics[width=0.32\textwidth,height=!]{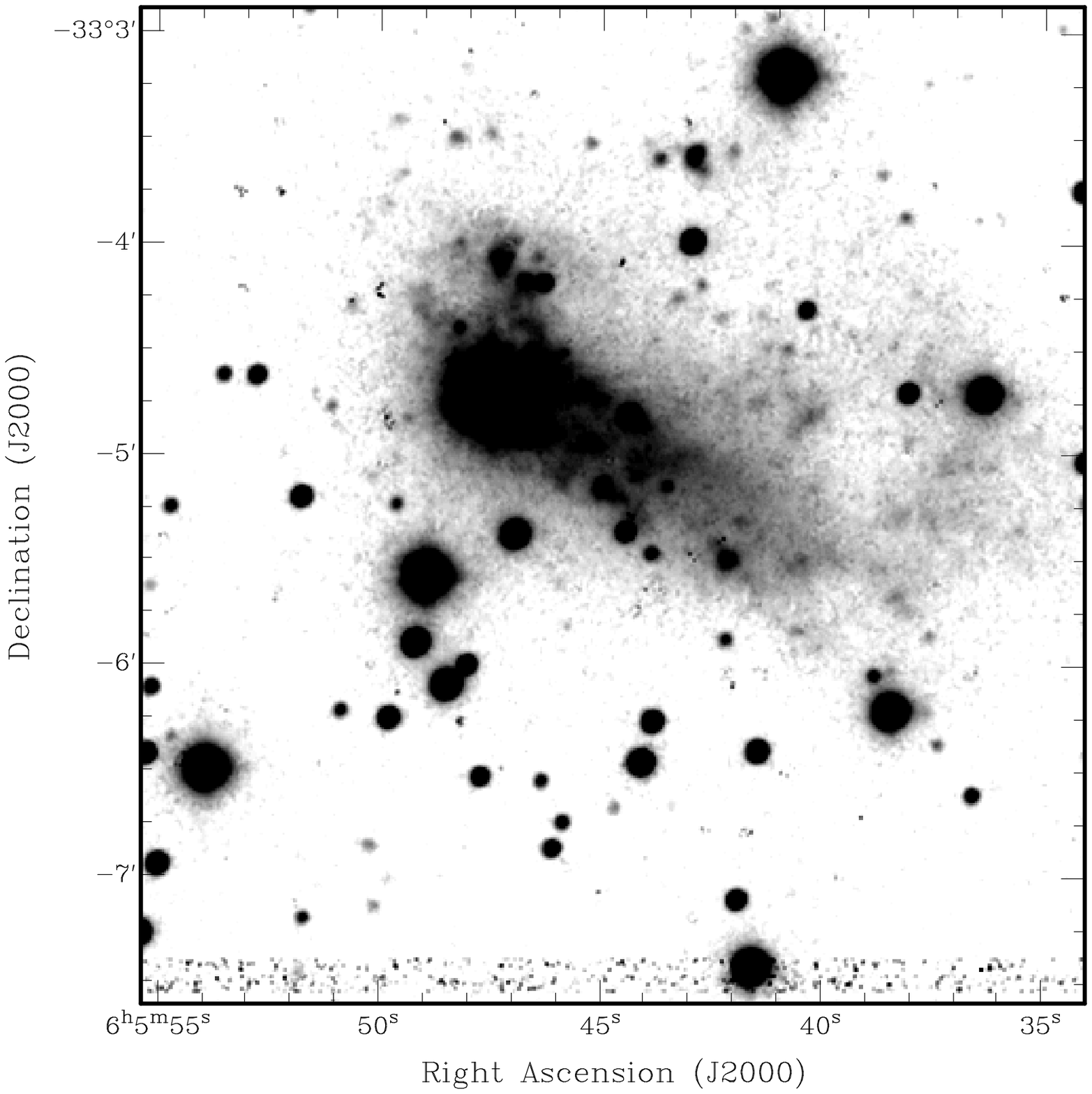} &
   \includegraphics[width=0.32\textwidth,height=!]{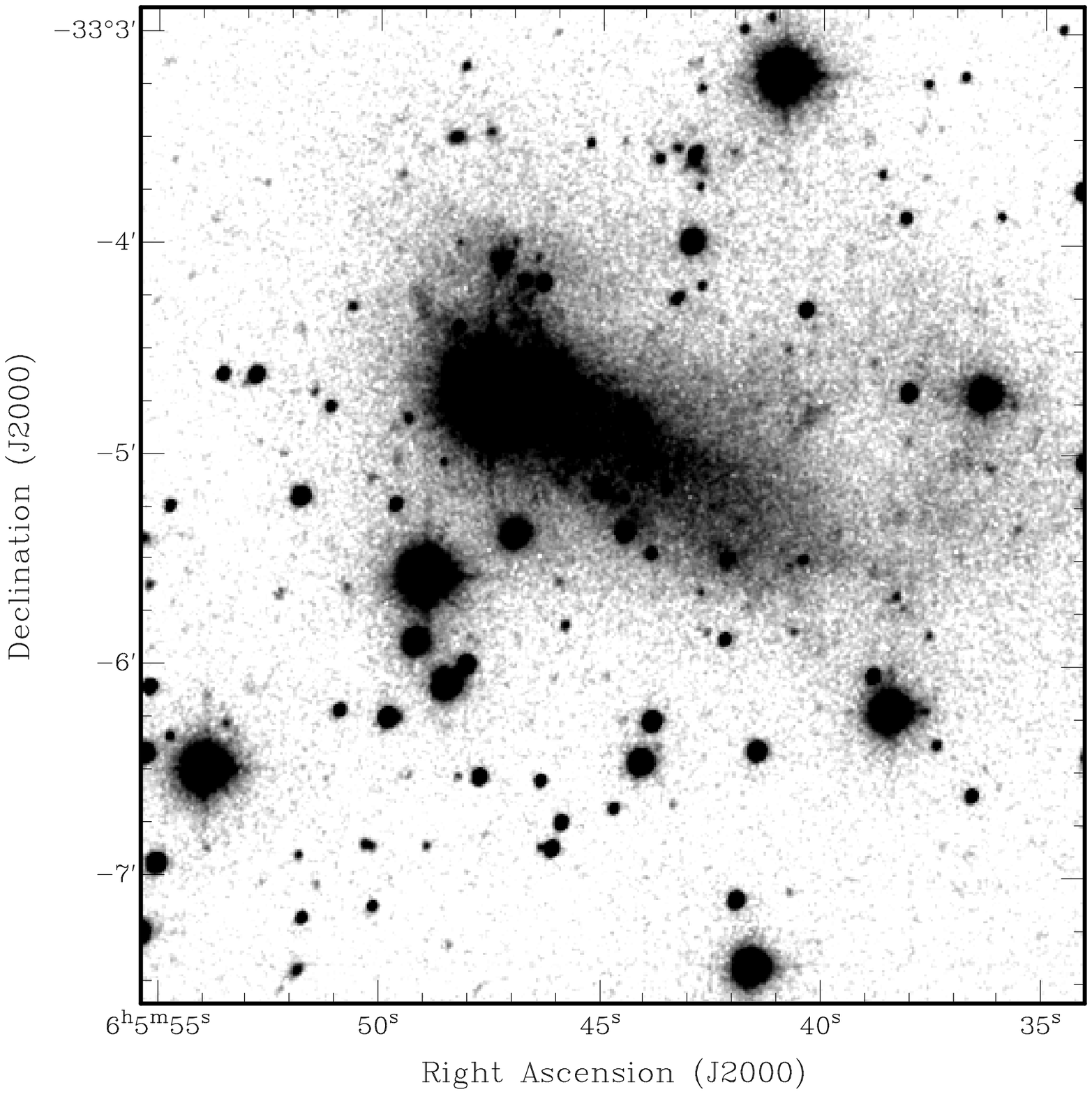} &
   \includegraphics[width=0.32\textwidth,height=!]{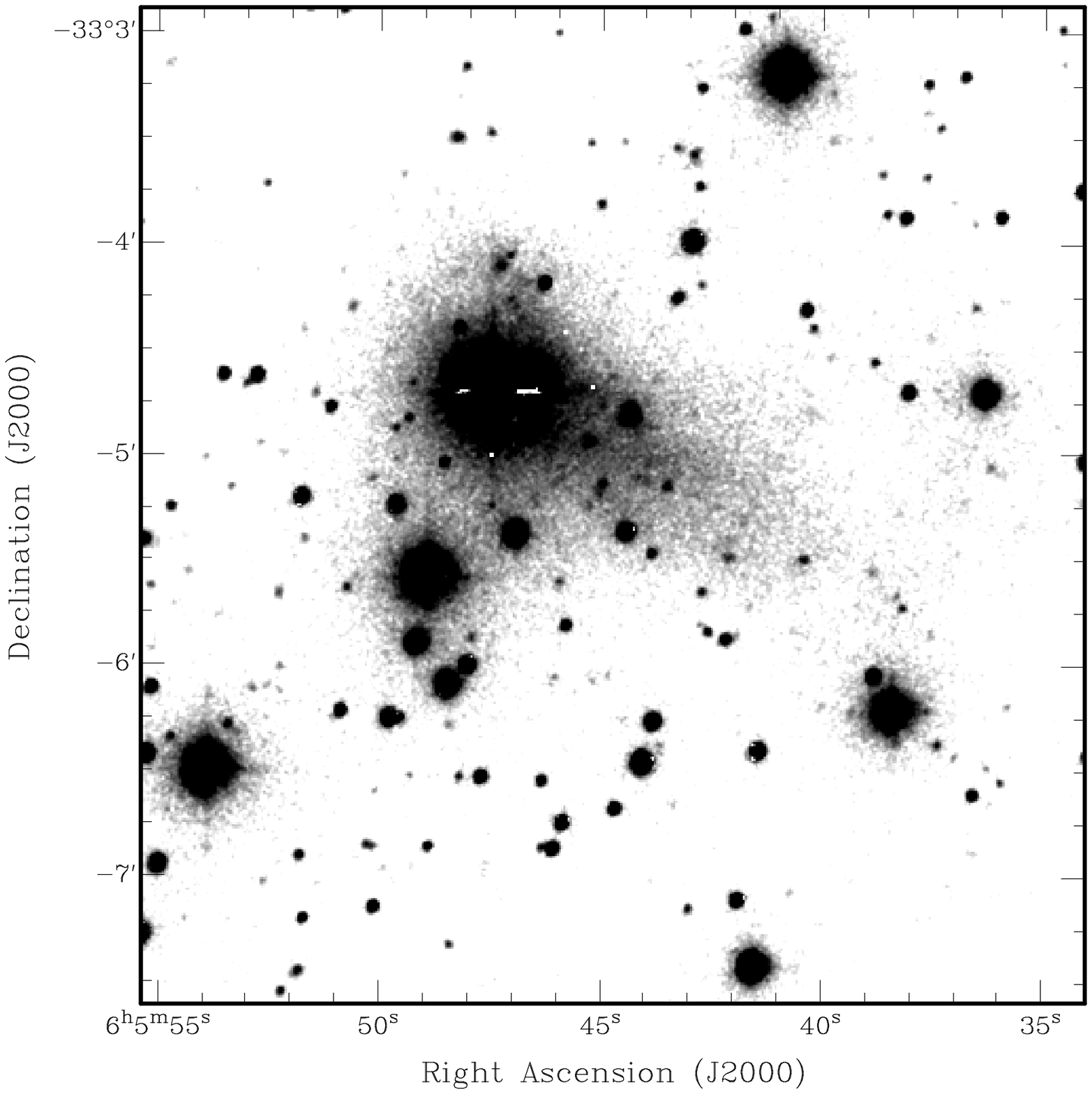} \\
 \end{tabular}
 \caption{Logarithmic grayscale images of \eso{} in the
   Cousins $B$ ({\em left}), $V$ ({\em middle}) and $I$ ({\em right})
   filters.The central bar and one spiral arm characterize \eso{} as a
Magellanic-type dwarf galaxy.  }
 \label{fig:opt_images}
\end{figure*}
\section{Optical Data\label{sec:optical}}
%
%
\subsection{Observations and Data Reduction\label{sec:opt_datared}}

Optical imaging observations were obtained on the nights of 9--11 April 1997
using
the 1-m telescope of Siding Spring Observatory (SSO) in Australia. A
total of $30$ broadband images combining short and long exposures were
obtained in the $B$, $V$ and $I$ Cousins filter bands. A $2048\times2048$
thinned TEK CCD with 15~$\mu$m pixels was used in direct imaging,
yielding a spatial sampling of $0\farcs61$~pix$^{-1}$. A
$1644\times1000$ CCD subsection with $100$~pixels virtual overscan was
read, for a total field-of-view of $15\farcm5\times10\farcm0$. The
seeing was moderate ($1\farcs7$--$2\farcs4$) but sufficient for basic
photometry and comparison with our \hi{} observations.

The data reduction was carried out using standard procedures in {\tt
IRAF}\footnote{{\tt IRAF} is distributed by the National Optical
Astronomy Observatories (NOAO), which are operated by the Association
of Universities for Research in Astronomy (AURA), Inc., under
cooperative agreement with the National Science Foundation (NSF).}
(\citealt{t86,t93}). All the images were overscan-subtracted,
bias-subtracted using bias scans obtained every night, and flatfielded
using twilight flatfields.
Dark current corrections were not necessary due to the low dark
current levels and the relatively short exposure times.
The cosmetic of the CCD is good, with only one bad column
and few bad pixels. All target images in a given filter were then
registered and combined using a pixel rejection algorithm, convolving
all images to a common (largest) seeing in each filter. Because
\eso{} is faint but lies in a region with many bright
foreground stars, all pixels with counts above $20\%$ of the
saturation level were rejected in the long exposure images, to prevent
excessive bleeding. Short exposures fill in the rejected regions,
easing the masking and interpolation of bright foreground stars (see
below).

Three sets of carefully selected and corrected Landolt $UBVRI$
standards (\citealt{l92,b95}) were used for photometric
calibration.
Since we do not require precise photometry, we adopted
the averaged extinction coefficients for the SSO 1-m telescope.
The resulting combined and
calibrated images are shown in Figure~\ref{fig:opt_images}.

%
%
\subsection{Surface Photometry} \label{sec:surf_phot}

Standard ellipse fitting using the {\tt Ellipse} and {\tt Isophote}
packages in {\tt IRAF} (see \citealt{j87}) was used to derive the
azimuthally-averaged surface brightness profile of
\eso{} in each filter. For each semi-major axis $a$,
the ellipses are characterized by the central coordinates
($x_0$,$y_0$), the ellipticity $\epsilon$
(defined as $\epsilon \equiv 1-b/a$, where $a$ and $b$ are the semi-major
and semi-minor axis, respectively) and the position angle PA, measured from
North to East.

The derived surface brightness profiles should be used with caution,
for the following reasons.
First, \eso{}'s irregular and lopsided morphology renders any
azimuthally-averaged profile at best a modest representation of its
two-dimensional light distribution.
Second, many bright foreground stars are present in the field of
\eso{}. Particularly troublesome is the bright
star \object{TYC\,7075-383-1} located near the North-East edge of the
central
bar at ($\alpha=06^{\rm h}05^{\rm m}47\fs6$,
$\delta=-33\degr04\arcmin41\farcs5$; J2000),
which has $B=12.1\pm 0.1$ and $V=11.4\pm 0.1$~mag (\citealt{k01}).
The most elegant way to remove the foreground stars is to subtract
the stellar profiles from the image.
However, the brightest stars in the field are saturated and
show bleeding effects even in our shortest exposures,
so we masked out the brightest foreground stars, and interpolated
the surface brightness in the masked regions.
As the images in each of the three filters have a comparable resolution, the
same
regions were used for all filters.

\begin{figure}[tb]
 \resizebox{0.9\hsize}{!}{\includegraphics{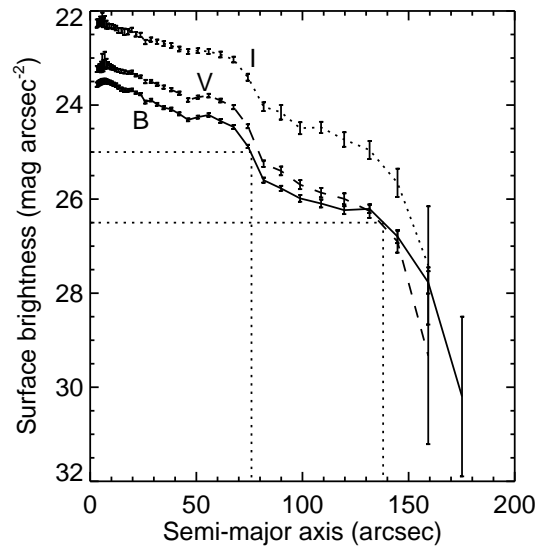}}
 \caption{The azimuthally-averaged surface brightness profiles of
   \eso{} in the $B$ band ({\em solid curve}), the $V$ band ({\em dashed
     curve}) and the $I$ band ({\em dotted curve}).
     The error bars indicate the $1\sigma$ formal errors.
   Systematic errors due to the irregular morphology of \eso{} are probably
much larger.
   The dotted lines indicate the semi-major axis of \eso{}
   at the $B=25$ and $26.5$~mag\,arcsec$^{-2}$ isophotes, respectively:
   $\rtf = 1\farcm27 \pm 0\farcm03$ and $\rhol = 2\farcm30 \pm 0\farcm08$. }
 \label{fig:opt_sbp}
\end{figure}

\begin{figure}[tb]
 \resizebox{0.9\hsize}{!}{\includegraphics{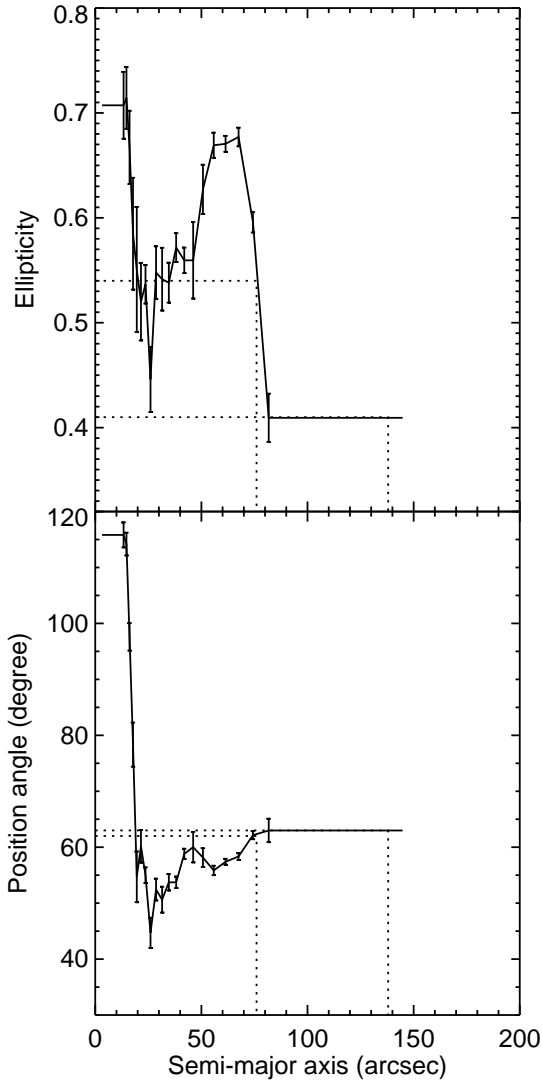}}
 \caption{Ellipticity $\epsilon$ ({\em top}) and position angle PA ({\em
bottom})
   profiles of \eso{}, derived with the azimuthally-averaged
   $B$ surface brightness profile shown in
   Figure~\ref{fig:opt_sbp}.
   Due to the irregular morphology of \eso{} it was not possible to
   successfully fit ellipses at radii larger than $83\arcsec$.
   $\epsilon$ and PA are assumed
   to be constant for radii larger than this value.
   The error bars indicate $1\sigma$ formal errors.
   Systematic errors due to the irregular morphology of \eso{} are probably
much larger.
   The dotted lines in both panels indicate the ellipticity and position
   angle at radii $\rtf$ and $\rhol$.
   }
 \label{fig:opt_epa}
\end{figure}

\begin{figure}[tb]
 \resizebox{0.9\hsize}{!}{\includegraphics{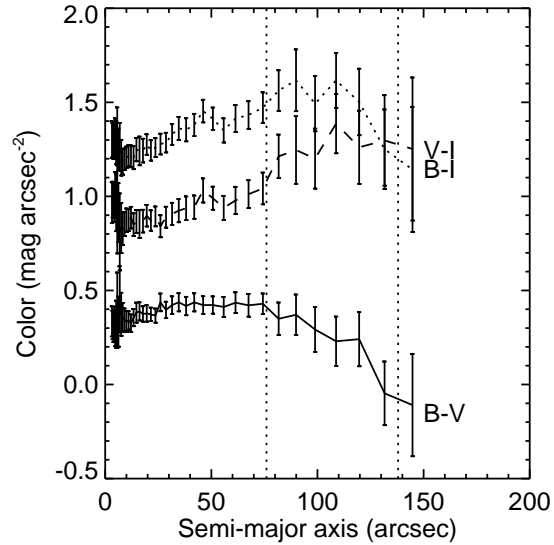}}
 \caption{Color profiles of \eso{}, derived from the
   azimuthally-averaged surface brightness profiles shown in
   Figure~\ref{fig:opt_sbp}: $B-V$ ({\em solid curve}), $V-I$ ({\em dashed
   curve}) and $B-I$ ({\em dotted curve}). The error bars indicate the
$1\sigma$ formal errors.
   Systematic errors due to the irregular morphology of \eso{} are probably
much larger.
   Vertical dotted lines are plotted at radii $\rtf$ and $\rhol$.}
 \label{fig:opt_color}
\end{figure}

Mainly because of \eso{}'s irregular morphology, it was difficult to
obtain reliable and stable fits for the whole range of possible semi-major
axes
($3\arcsec$ to $155\arcsec$). Our results are shown in
Figures~\ref{fig:opt_sbp}--\ref{fig:opt_color}.
The $I$-band images are less deep because of the relatively bright
background.
In order to get meaningful colors, the ellipses were fitted on
the $B$ image only and imposed on the $V$ and $I$ band images. One can see
that the
ellipticity and position angle of the ellipses were successfully
fitted over a limited range of semi-major axes only, $15\arcsec\la a
\la83\arcsec$. For lack of a better prescription, they are assumed
constant outside of those limits.


The light profiles are approximately exponential over most of the
optical extent ($a \la 135\arcsec$), falling off
with roughly 1.38 mag\,arcmin$^{-1}$.
A break to a steeper fall-off is seen at $a\approx135\arcsec$.
A fit to the $B$ profile within that region
yields a radial scale length $h_B=50\arcsec \pm 5\arcsec$ and an
(uncorrected)
extrapolated central surface brightness $\mu_B(0)=23.3$~mag~${\rm
arcsec}^{-2}$. The standard criterion for defining low surface
brightness galaxies is for the extrapolated central surface brightness
(based on the outer exponential disk) to be equal or fainter than
$23.0$~mag~${\rm arcsec}^{-2}$ in $B$ (\citealt{impey1997}).
Following this definition, \eso{} would qualify as a low surface brightness
galaxy.
The $B$ surface brightness profile also suggests a radius at
the $25$~mag~${\rm arcsec}^{-2}$ isophote of $\rtf=1\farcm27 \pm 0.03$,
identical to the values listed in HyperLEDA\footnote{HyperLEDA:
http://leda.univ-lyon1.fr~.} and the \citeauthor{vetal91}. The
Holmberg radius (defined at the $26.5$~mag~${\rm arcsec}^{-2}$
isophote in $B$; see \citealt{holmberg1958}) is $\rhol=2\farcm30 \pm
0\farcm08$.

Over most of the disk $B-V =0.4 \pm 0.08$~mag, slightly bluer in the very
inner and outer parts. This is marginally bluer than the \lmc{}
($B-V=0.51$~mag; HyperLEDA) but similar to \object{NGC~4618} (SBm;
\citealt{o91}). The $V-I$ color increases from $0.85 \pm 0.08$~mag in the
inner parts
to $1.3 \pm 0.1$~mag in the outer parts, again similar to \object{NGC~4618}.

Constructing growth curves from the surface brightness profiles shown
in Figure~\ref{fig:opt_sbp}, the extrapolated total apparent magnitudes
are $B_{\rm T}=13.8 \pm 0.1$, $V_{\rm T}=13.5 \pm 0.1$ and $I_{\rm T} = 12.4
\pm 0.15$~mag, in good
agreement with the values listed in the \citeauthor{vetal91}
($B_{\rm T}=13.58\pm0.21$~mag), HyperLEDA ($B_{\rm T}=13.81\pm0.22$~mag) and
The Surface
Photometry Catalogue of the ESO-Uppsala Galaxies ($B_{\rm
T}=13.6\pm0.1$~mag;
\citealt{lv89}). Correcting for Galactic extinction following
\citet{sfd98}, but neglecting the inclination-dependent internal
absorption, we obtain the following corrected total apparent magnitudes:
$B_{\rm c}\approx13.6\pm 0.1$, $V_{\rm c}\approx13.4$ and $I_{\rm
c}\approx12.3$~mag.
%
%
\section{\hi{} Radio Synthesis Data\label{sec:hi}}
%
%
\subsection{Observations and Data Reduction\label{sec:hi_datared}}

Radio synthesis data of \eso{} were obtained with four
different configurations of the Australia Telescope Compact Array
(ATCA). Full $12$~hour syntheses were obtained with the 375, 750C and
750E arrays, while shorter observations ($3$--$4$~hour on-source) were
also obtained twice in the 1.5D array. Antenna CA06, providing the
longest possible baselines, was not used in the 375, 750C and 750E
arrays. Both \hi{} line and continuum data (centered at
$1380$~MHz) were obtained simultaneously in two polarizations, but
only the line data will be discussed here. In the 375, 750C and 750E
configurations, $512$~channels in each polarization were used, each
$1.65$~\kms{} wide, for a total bandwidth of
$844$~\kms{}. In the 1.5D configuration, $256$~channels of
$6.60$~\kms{} width were used in each polarization for a total
coverage of $1689$~\kms{}. These and other relevant observing
parameters are listed in Table~\ref{tab:radio_par}.

\begin{table}[tb]
 \caption{Parameters of the ATCA \hi{} observations. Observations were
obtained in four different configurations, and combined (see
\S~\ref{tab:radio_datasets}) for analysis.}
 \label{tab:radio_par}
 \begin{tabular}{p{2.2cm}p{2.5cm}p{3cm}}
   \hline
   \hline
   ATCA~array              & Date of obs.      & Maximum baseline \\
   \hline
   375                      & 1997 Oct~3--4    & $459$~m\\
   750C                     & 1997 Oct~25--26  & $750$~m\\
   750E                     & 1998 May~30      & $643$~m\\
   1.5D                     & 1997 Mar~8, 10   & $3000$~m\\
   \hline
   \hline
   \\
   \hline
   \hline
   Property                        & 1.5D array                 & 375, 750C,
750E arrays       \\
   \hline
   Primary~beam                    & HPBW$\approx33\arcmin$     &
HPBW$\approx33\arcmin$     \\
   Correlator                      & FULL\_8\_512-128           &
FULL\_4\_1024-128          \\
   Channel width                   & $31.3$~kHz                 & $7.8$~kHz
\\
                                   & ($6.60$~\kms{})            &
($1.65$~\kms{})            \\
   Bandwidth                       & $8.00$~MHz                 & $4.00$~MHz
\\
                                   & ($1689$~\kms{})            &
($844$~\kms{})             \\
   Flux~calibrator                 & \object{PKS~1934-638}               &
\object{PKS~1934-638}               \\
   Phase~calibrator                & \object{PKS~0614-349}               &
\object{PKS~0614-349}               \\
   \hline
   \hline
 \end{tabular}\\
 Note: Antenna CA06 was not used with the 375, 750C and 750E arrays.\\
\end{table}

The data were reduced using standard procedures in {\tt
MIRIAD}\footnote{{\tt MIRIAD} was developed by the Australia Telescope
National Facility (ATNF) which is part of the Commonwealth Scientific
and Industrial Research Organization (CSIRO) of Australia.}
(\citealt{stw95}) and visualized with {\tt KARMA} (\citealt{g95}). The
strong source \object{PKS~1934-638} (\citealt{reynolds1994}) was observed once
briefly in each
configuration for use as a flux calibrator. \object{PKS~0614-349},
located $1\degr26\arcmin$ to the South-East of
\eso{}, was observed roughly every $50$~minutes for use as
a complex gain and bandpass calibrator.

The (ATCA-specific) self-interference channels were first removed, after
which
the line data were transformed to heliocentric rest frame and Hanning
smoothed. Bad data, mostly due to short duration interference, was
then masked out manually and the flux density, complex gains, and bandpass
were calibrated. The continuum emission was subtracted in the $uv$ plane
using a linear fit to line-free channels on each side of
\eso{}'s emission (excluding bandpass-affected
channels), and the $uv$ data were then imaged to create dirty
cubes.

Both uniform- and natural-weighted cubes were created and
a $33\arcmin\times33\arcmin$ box  (equal to the primary beam
diameter) was imaged. In each case the pixel size was chosen to sample the
synthesized beam with 2 to 3~pixels.
The four datacubes used in our analysis are listed in
Table~\ref{tab:radio_datasets},
and are hereafter abbreviated as NH (naturally-weighted, high-sensitivity
datacube),
UH (uniformly weighted, high-sensitivity datacube), NL (naturally weighted,
low-sensitivity
datacube) and UL (uniformly weighted, low-sensitivity datacube).
For datacubes UH and NH we merge the 375, 750C and 750E arrays only, while
for
datacubes UL and NL we combine the data for all arrays.
Note that the low-sensitivity datasets (NL and UL) have a significantly
higher spatial resolution. In the analysis below we use the
the high-sensitivity datasets for global properties
and the high-resolution datasets for the small-scale morphology.
Table~\ref{tab:radio_datasets} lists
the root-mean-square (rms) noise
per channel and synthesized beam for each dataset.
When considering the noise levels reached, one
should remember that the channel width for the 1.5D observations was
four times as large as that for the other arrays, although these
observations were regridded onto the narrower $1.65$~\kms{}
channels when combined with the shorter arrays.

\begin{table*}[tb]
 \caption{Properties of the four datacubes used in our analysis. For the
analysis of the detailed morphology of \eso{} we use the long-baseline
datacubes NL and UL, which have a high spatial resolution. For the
derivation of the global structure of \eso{} we use the datacubes~NH and UH,
which contain only the short baseline observations. The latter datasets are
therefore deeper than the former, but have lower spatial resolutions. }
 \label{tab:radio_datasets}
 \begin{tabular}{lllclccr}
   \hline
   \hline
   Datacube & Properties & Arrays used & Pixel    & Weighting         & rms
noise  & \multicolumn{2}{l}{Beam properties}  \\
    &                  & & size       & & per channel       & FWHM       &
\multicolumn{1}{c}{$\varphi$}    \\
    &                  & & (arcsec)   & & (mJy beam$^{-1}$) & (arcsec)   &
\multicolumn{1}{c}{(deg)} \\
   \hline
   Datacube NH & high sensitivity, low resolution & 375, 750C and 750E &
$16\times16$ & natural & $2.0$ & $132\times70$    & $ -1$ \\
   Datacube UH & high sensitivity, low resolution  & 375, 750C and 750E &
$16\times16$ & uniform & $3.3$ &  $72\times45$    & $  0$ \\
   Datacube NL & low sensitivity, high resolution & All arrays        &
$4\times4$ & natural & $1.5$ &  $58\times38$    & $ +8$ \\
   Datacube UL & low sensitivity, high resolution & All arrays        &
$4\times4$ & uniform & $2.6$ &  $21\times10$    & $+30$ \\
   \hline
 \end{tabular}
\end{table*}

\begin{figure*}[tb]
 \centering
 \includegraphics[width=17cm]{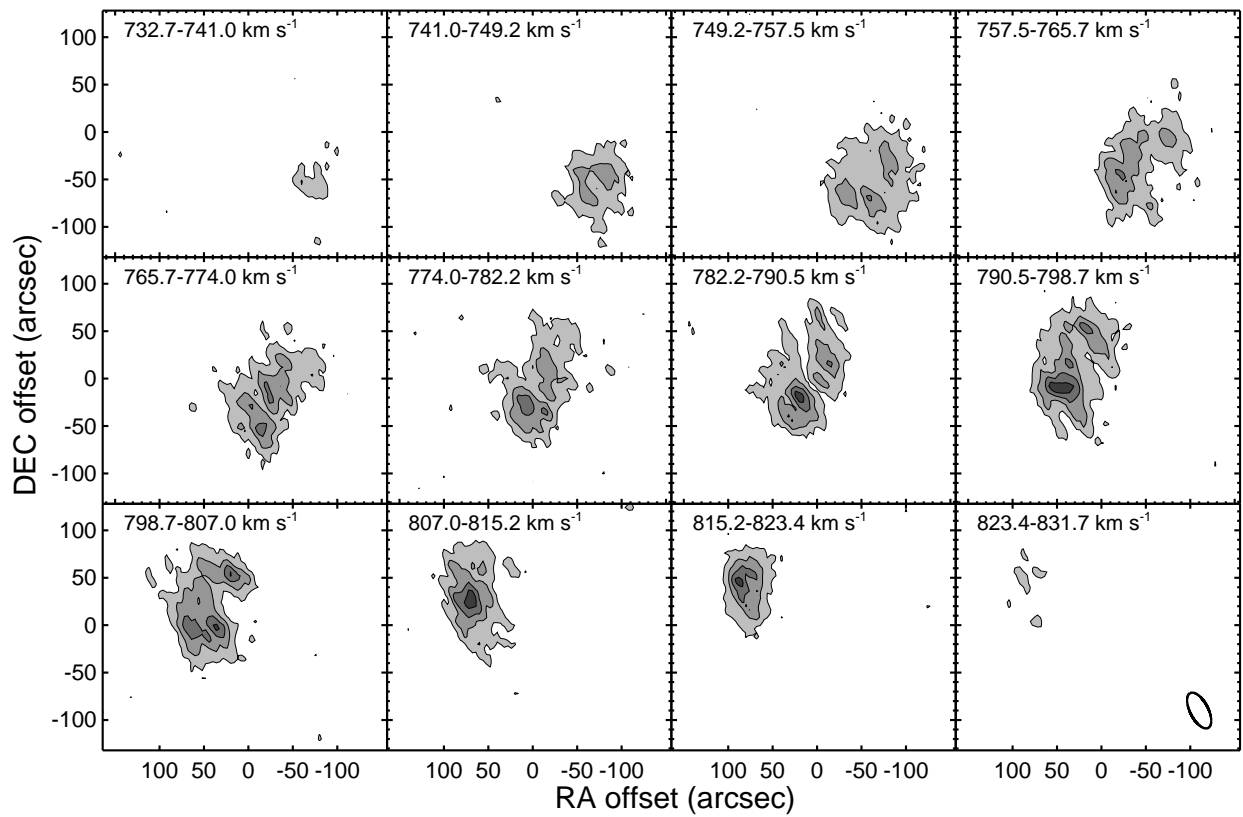}
 \caption{\hi{} channel maps of \eso{} for datacube~UL
 (see Table~\ref{tab:radio_datasets}).
   Each map is the average of
   $5$ channels to increase the sensitivity ($8.25$~\kms{} per
   map). Contour levels are at $3$, $6$, $9$ and $12$ times the rms
   noise in each map ($1.2$~mJy~beam). The
   heliocentric velocity range of each map is indicated in the top-left
   corner. The synthesized beam is $21\arcsec\times10\arcsec$ with a
   position angle of $30\degr$, and is indicated in the bottom-right panel. }
 \label{fig:chan_maps}
\end{figure*}

The Clark Clean algorithm (\citealt{c80}) was used for the
deconvolution, cleaning an area tightly encompassing the emission in
each channel. The maps were cleaned until the total cleaned flux
stopped increasing, typically at a depth of $2.5$ times the rms noise
per channel, which was sufficient to remove all sidelobes. The clean
components were then restored and added to the residuals using a
Gaussian clean beam as listed in Table~\ref{tab:radio_datasets} (and
resulting from a fit to the inner parts of the dirty beam). Channel
maps of datacube~UL (with the best spatial resolution)
are shown in Figure~\ref{fig:chan_maps}, where each
map displayed is the average of $5$ channels to increase
sensitivity.

%

%
%
\subsection{Global \hi{} Profile\label{sec:glob_hi}}

\begin{figure}[tb]
  \includegraphics[width=0.5\textwidth,height=!]{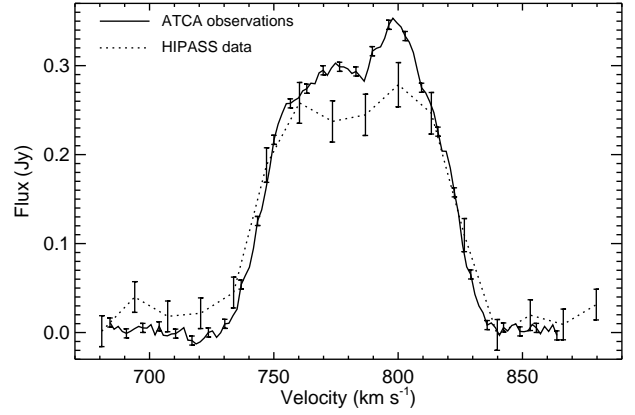}
    \caption{The global \hi{} profile of \eso{}
     ({\em solid line}), obtained from
   datacube~NH by integrating the individual channel maps over the area
   containing emission. The total \hi{} flux is $F_\hi=23.1\pm
1.2$~\jykms{}.
   The HIPASS global profile summed over
   a $40\arcmin\times40\arcmin$ area is overplotted ({\em dotted line}).
   For clarity, the error bars for our ATCA data are only plotted for each
fourth channel.
   }
 \label{fig:hi_glob_prof}
\end{figure}

To maximize the flux recovered, the (masked) NH~datacube
was used to derive the global profile of
\eso{}, which is shown in Figure~\ref{fig:hi_glob_prof}. The
linewidth (uncorrected for instrumental broadening) at $20\%$ of the
peak is $\Delta V^{\rm obs}_{20}=89 \pm 2$~\kms{}, while that at $50\%$
$\Delta V^{\rm obs}_{50}=73 \pm 2$~\kms{}. We adopt a systemic velocity,
taken
to be the midpoint of the velocities at $20\%$ of the peak, of $V_{\rm
sys}=784 \pm 2$~\kms{}. An intensity-weighted mean of the entire
profile gives $\langle V \rangle=785 \pm 2$~\kms{},
consistent with the former value.
\citet{lhgmw82} obtained $\langle V\rangle=790\pm5$~\kms{} and $\Delta
V_{50}=85\pm10$~\kms{}
from Parkes Telescope data (HPBW$=15\arcmin$). Both values are in good
agreement
with our observations, especially considering that the slightly lower
spectral resolution
of \citet{lhgmw82} may have led to a mild overestimation of $\Delta V_{50}$.
\citet{hkk01} list $\langle V \rangle=787\pm3$~\kms{} and
$\Delta V_{50}=65$~\kms{} from HIPASS, also using Parkes.

After integration over the full profile in Figure~\ref{fig:hi_glob_prof}, we
find a total \hi{} flux $F_\hi=23.1\pm1.2$~\jykms{}.
\citet{lhgmw82} obtained
$36\pm4$~\jykms{} but \citet{hkk01} list $13.7$~\jykms{},
while \citet{ketal04} and \citet{metal04} respectively quote
$17.6\pm2.5$ and $17.2$~\jykms{} from HIPASS data
(the latter HIPASS measurements were obtained under the assumption that
\eso{} is a point source).
It is thus unclear from those if our synthesis
observations are missing any short spacing flux or not. We reanalyzed
the HIPASS archive observations ourselves, summing the data in a
$40\arcmin\times40\arcmin$ box, and we find a total \hi{} flux of
$21.8\pm1.0$~\jykms{}, still lower than the result of
\citet{lhgmw82} but consistent with our own ATCA measurement. This
HIPASS global profile is also shown in Figure~\ref{fig:hi_glob_prof}.

Additional uncertainties in the \hi{} flux include the
possibility of \hi{} self-absorption and the existence of dense
gas clumps associated with the apparently diffuse \hi{} gas
(\citealt{grenier2005}). Given that $N_\hi/\tau \approx T_s~\Delta~v)$,
where $N_\hi$ is the column density, $\tau$ the opacity, $T_s$ the spin
temperature, and $\tau$ and $T_s$ are correlated (\citealt{dickey1990}),
it is not easy to calculate a definite upper limit for $N_\hi$
even from the line analysis. Simplifications can however be made using
the isophotal axis ratio as a measure of the disk inclination
(\citealt{giovanelli1994}). The correction factor, $(a/b)^{0.12}$, for
the disk of \eso{} is about $1.1$, so this effect is likely
small; we have not corrected for \hi{} self-absorption above.

The ATCA global profile shown in Figure~\ref{fig:hi_glob_prof} appears
asymmetric, with more \hi{} at higher velocities. This
behaviour is common among late-type dwarfs (e.g.\
\citealt{rs94,sahs02}). \cite{schoenmakers1997} have developed
a complex algorithm to quantify asymmetry using Fourier techniques.
In our analysis, however, we restrict ourselves to the 
standard moment maps, and 
quantify the importance of the asymmetry with the parameter
\begin{equation}
\label{eq:hi_asym}
Q_{\rm asym}\equiv\frac{\left|\,\int_{-\infty}^{V_{\rm
sys}}I(v)\,dv-\int_{V_{\rm sys}}^{\infty}I(v)\,dv\,\right|}
{\int_{-\infty}^{\infty}I(v)\,dv}=1.7\% \,,
\end{equation}
where $I(\nu)$ is the flux as a function of frequency $\nu$.
In the expression above, we adopted integration limits of $720$ and
$860$~\kms{}, respectively (as for the mean velocity and total
fluxes above). The global profile asymmetry of \eso{}
is thus mild and the HIPASS data show an even milder asymmetry.
%
%
\subsection{\hi{} Moments and Rotation Curve\label{sec:rot_cur}}

\begin{figure}[tb]
 \resizebox{0.9\hsize}{!}{\includegraphics{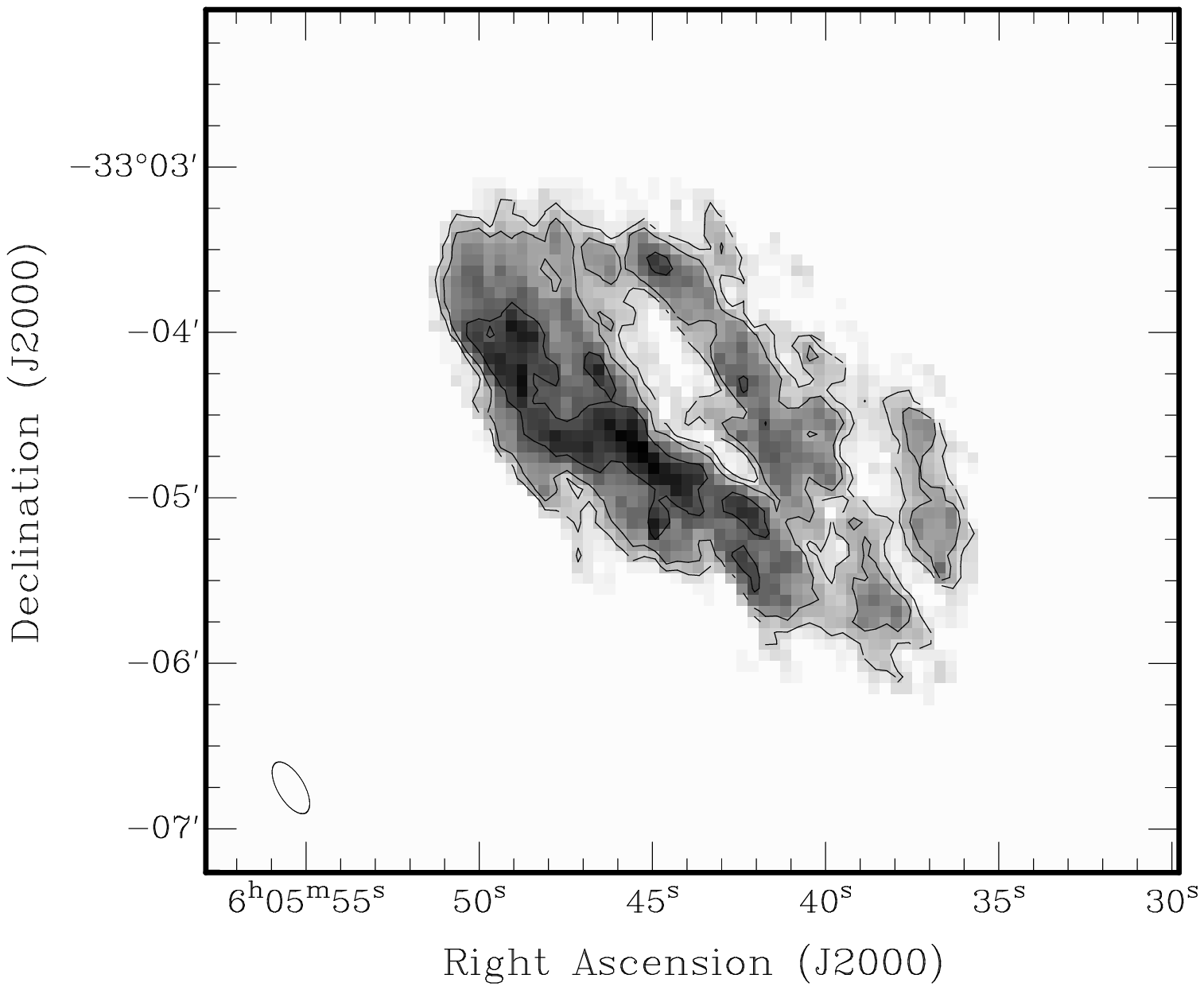}}
 \vspace*{0.25cm}\\
 \resizebox{0.9\hsize}{!}{\includegraphics{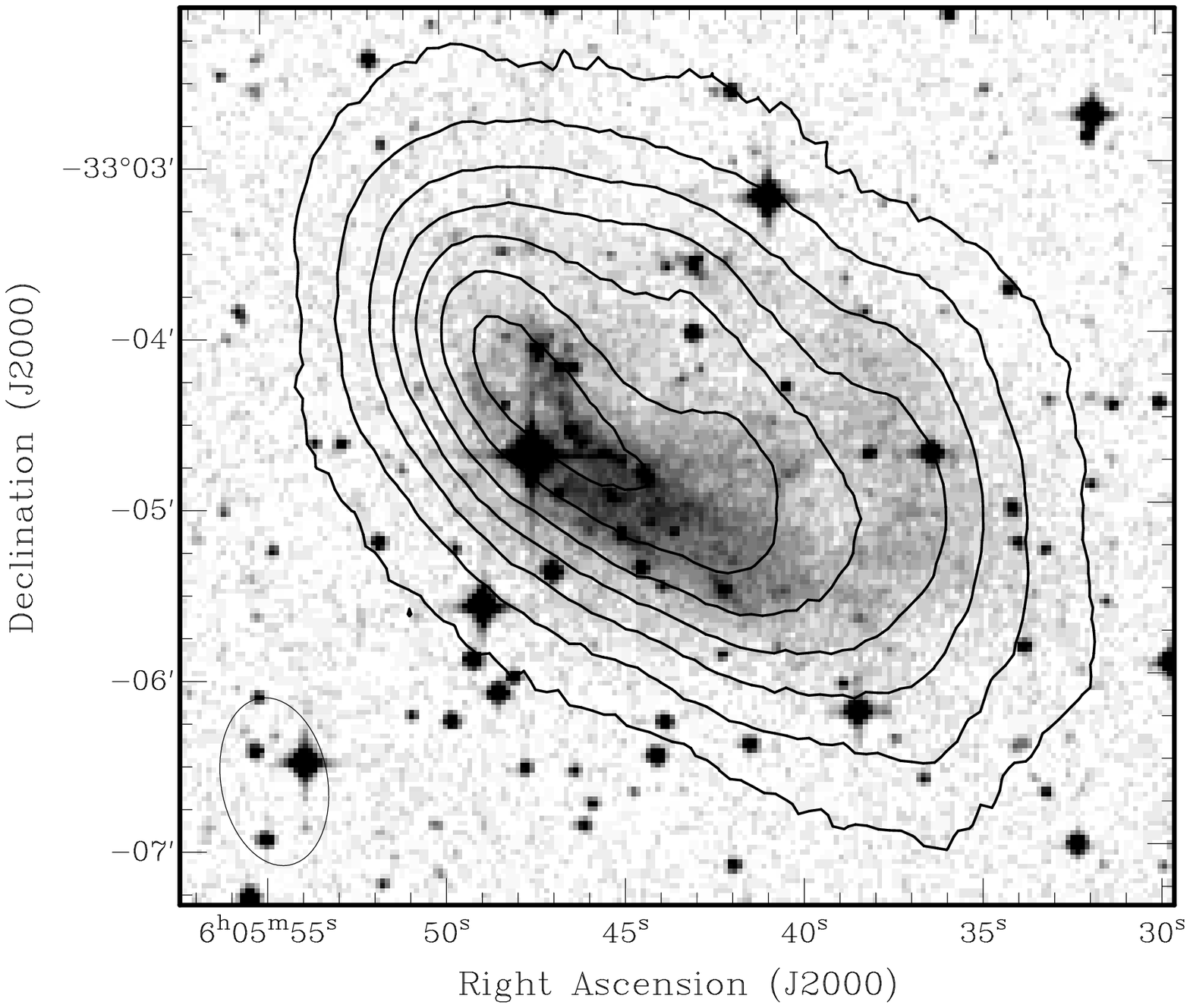}}
 \caption{Total \hi{} maps of \eso{} for
   the low sensitivity (high resolution) datacubes.
   {\em Top:} UL isodensity contours
   overlaid on a grayscale representation. The contours represent $1$,
   $2$ and $4$ times the faintest contour of
   $50$~mJy~beam$^{-1}$~\kms{} or $2.65\times10^{20}$~cm$^{-2}$.
   {\em Bottom:}
   NL contours overlaid on a DSS image. The
   contours represent $1$, $5$, $10$, $15$, $20$, $25$ and $30$ times the
   faintest contour of $80$~mJy~beam$^{-1}$~\kms{} or
   $0.4\times10^{20}$~cm$^{-2}$. The contours generally increase toward
   the center and the synthesized beam is indicated in the bottom-left
   corner of each map.
   }
 \label{fig:hi_mom0}
\end{figure}

\begin{figure}[tb]
 \resizebox{0.9\hsize}{!}{\includegraphics{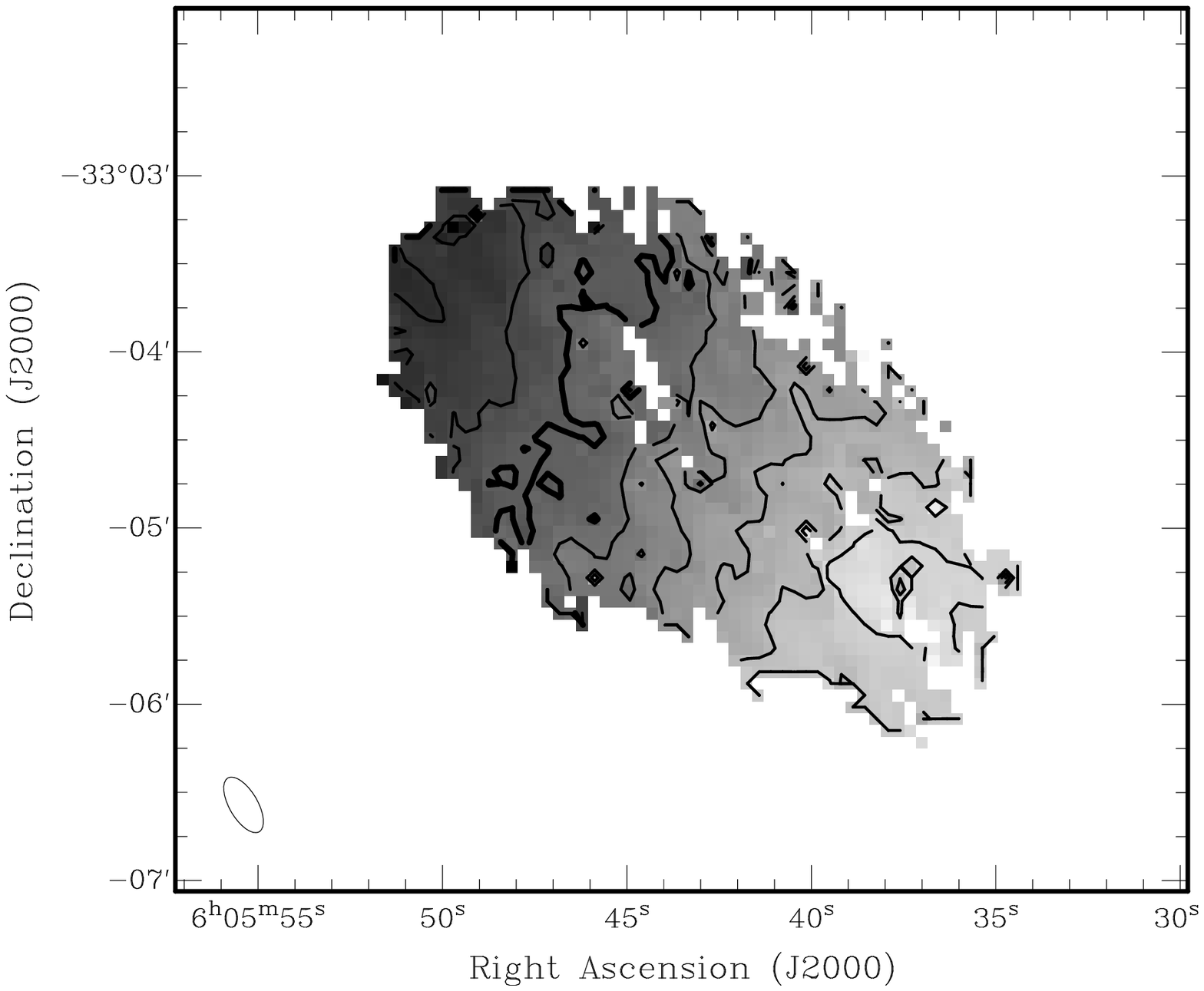}}
 \vspace*{0.25cm}\\
 \resizebox{0.9\hsize}{!}{\includegraphics{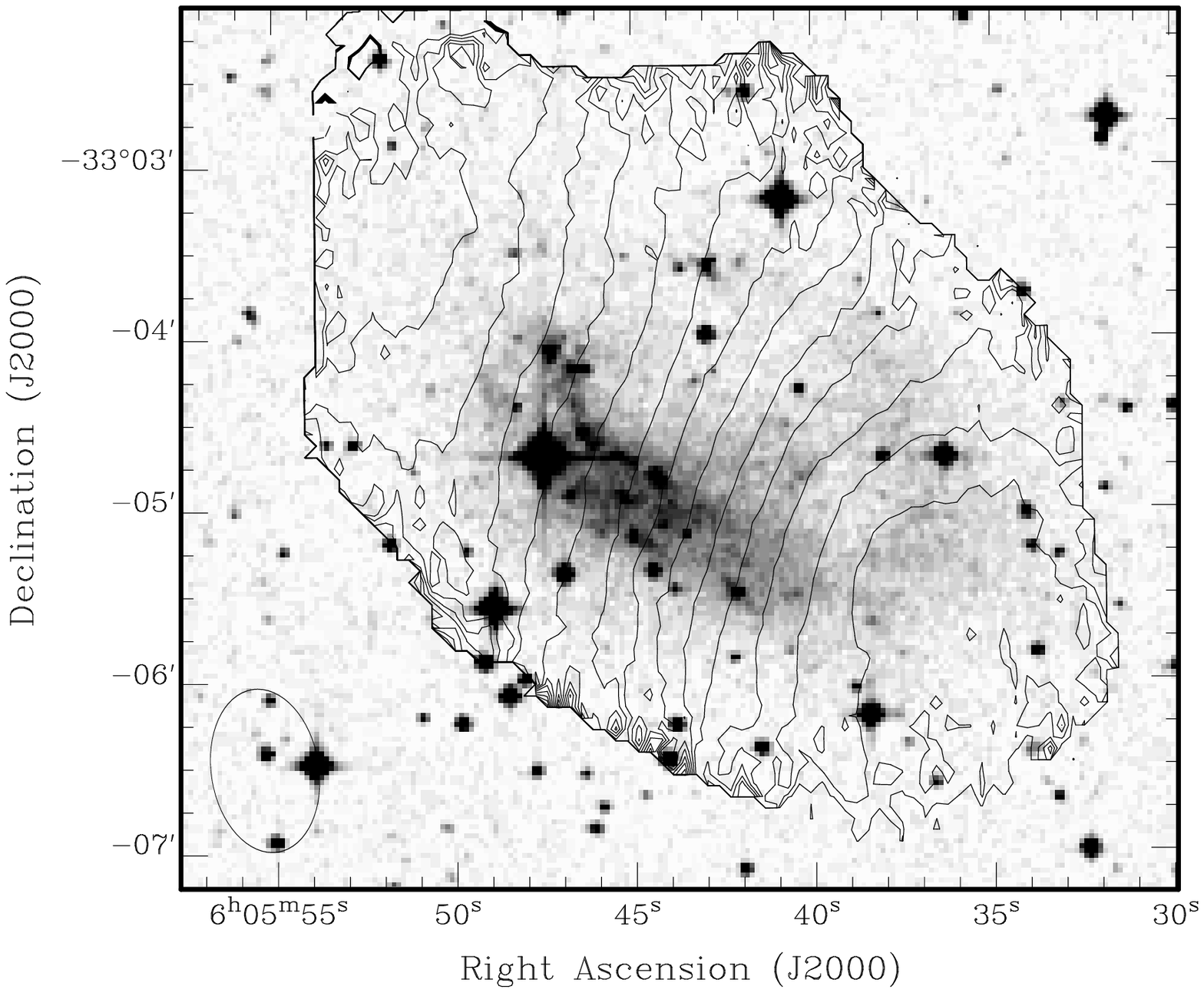}}
 \caption{Same as Figure~\ref{fig:hi_mom0} but for the velocity fields of
   \eso{}. The isovelocity contours increase toward the North-East
   and cover the velocity range $750$--$820$~\kms{} in both
   panels. The contours are spaced by $10$~\kms{} ({\em top}) and
   $5$~\kms{} ({\em bottom}).}
 \label{fig:hi_mom1}
\end{figure}

\begin{figure}[tb]
 \resizebox{0.9\hsize}{!}{\includegraphics{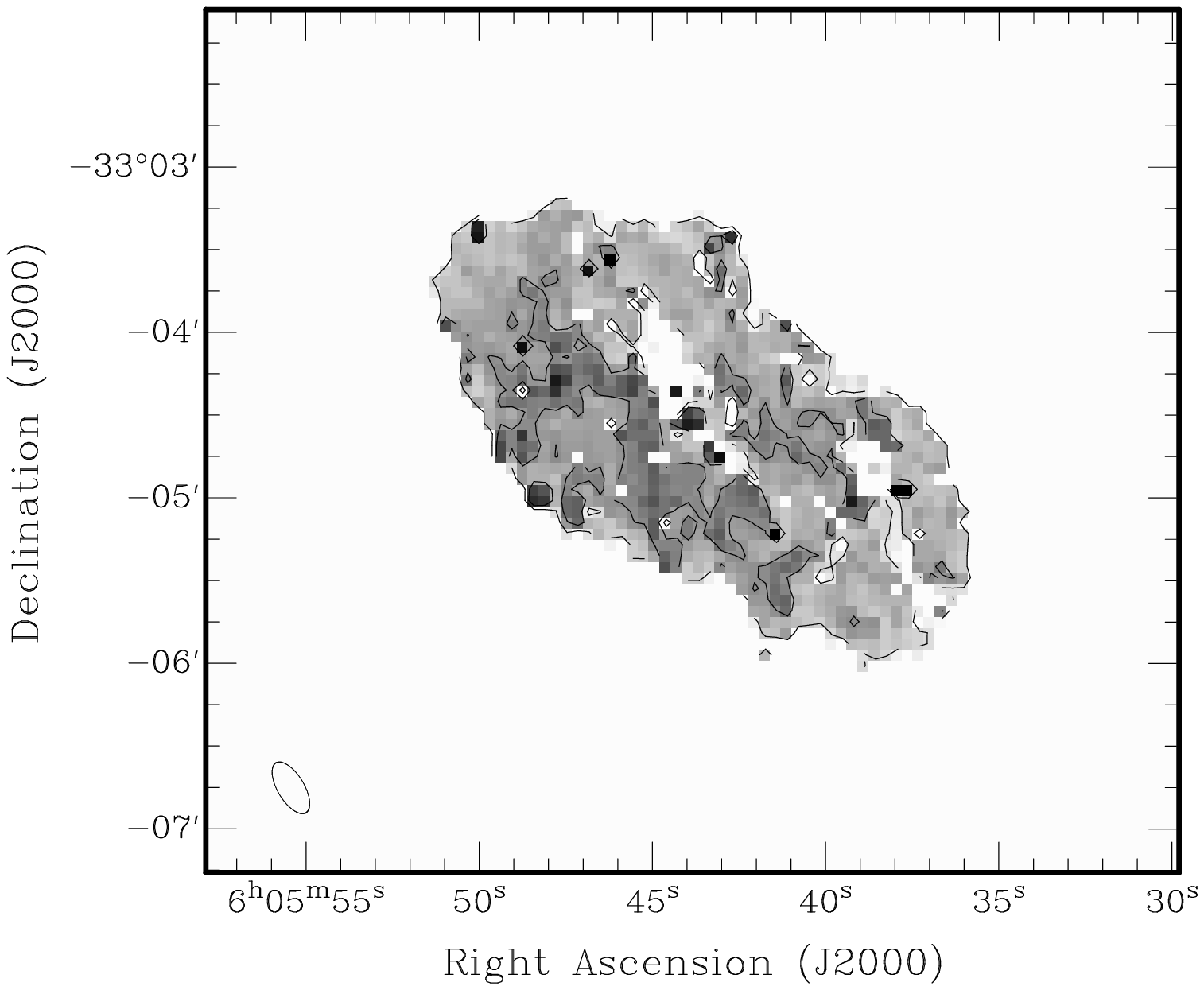}}
 \vspace*{0.25cm}\\
 \resizebox{0.9\hsize}{!}{\includegraphics{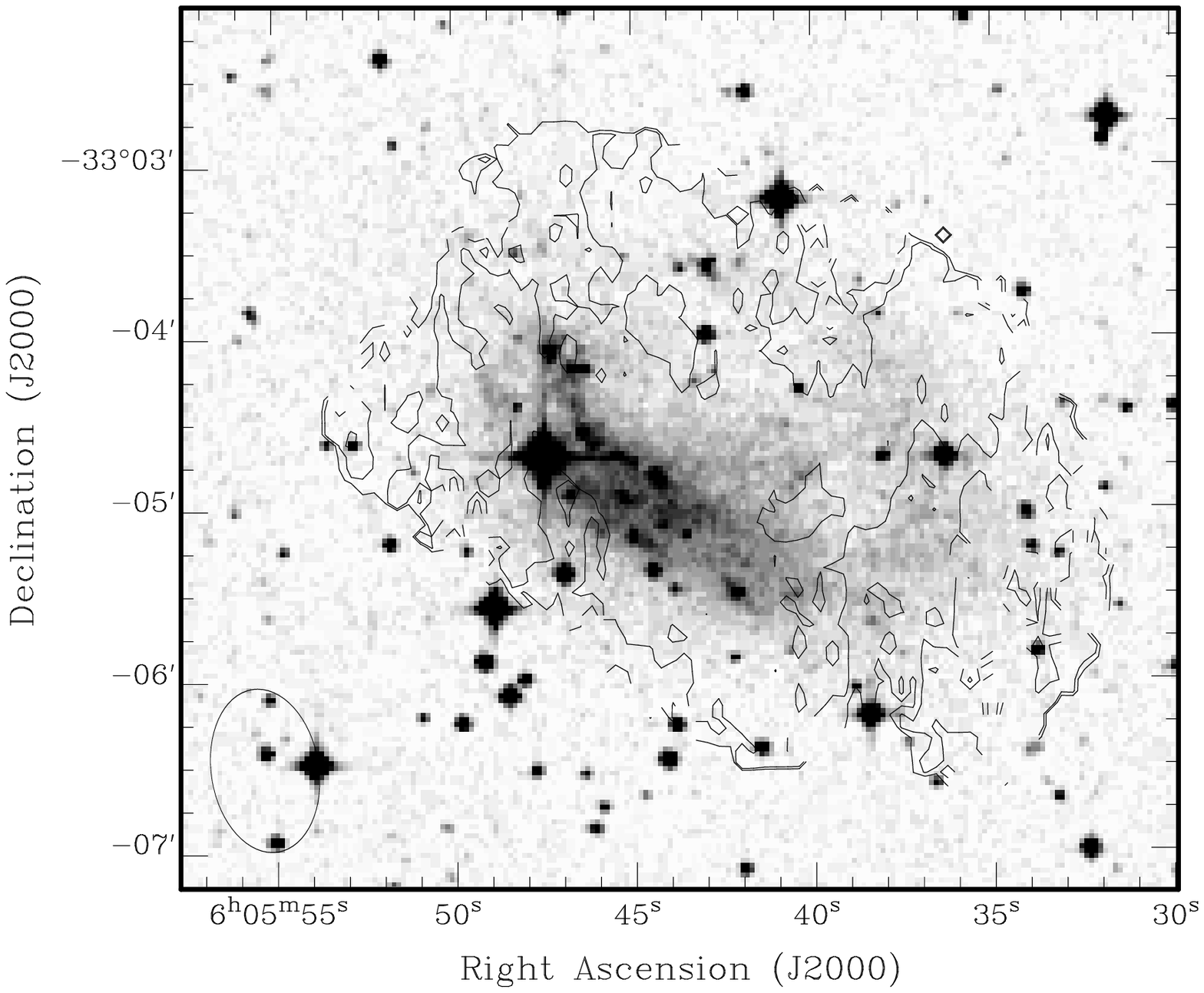}}
 \caption{Same as Figure~\ref{fig:hi_mom0} but for the velocity
   dispersion fields of \eso{}. The contours generally
   increase toward the center and represent $2$ and $6$~\kms{}
   ({\em top}) (unreliable) and $8$ and $12$~\kms{} ({\em bottom}).}
 \label{fig:hi_mom2}
\end{figure}

Moment maps were derived in the standard manner from the cleaned cubes
of the datasets listed in Table~\ref{tab:radio_datasets}.
Figures~\ref{fig:hi_mom0}--\ref{fig:hi_mom2} show the total
\hi{} map, the velocity field and the velocity dispersion
field for the high-resolution dataset, both uniform and
natural-weighted. All significant emission is contiguous so a mask was
derived from the zeroth moments and applied to the others.


The total \hi{} maps (Figure~\ref{fig:hi_mom0}) show that the
\hi{} roughly follows the optical light, although slightly offset
to the North-East, and has the same asymmetries. The map from datacube~UL
shows very well that the \hi{} is
concentrated along the main bar and the one-arm spiral, although there
is a third component intermediate between those two in terms of
position, length, and mass. While this third component has a stellar
counterpart where it joins with the bar, this does not appear to be
the case over its entire length. The velocity fields
(Figure~\ref{fig:hi_mom1}) reveal solid-body rotation over the entire
optical extent, a common feature among dwarf irregular galaxies (e.g.\
\citealt{bmh96,s99}). Interestingly, no kinematic lopsidedness is
present and the one-arm spiral has no obvious influence on the
kinematics, although this may simply be related to the poor spatial
resolution of most maps. The rotation curve appears to flatten at the
largest radii and a kink in the kinematic major-axis suggests the
presence of a warp in the outer parts. The velocity dispersion is
$8$--$16$~\kms{}
over most of the galaxy
(Figure~\ref{fig:hi_mom2}), the higher values being reached over the
optical bar and the third component.

The {\tt GIPSY}\footnote{Groningen Image Processing System.}
(\citealt{htbzr92,vt01}) implementation of \rotcur{}
(\citealt{b87}) was used for rotation curve fitting. The target was
divided into concentric tilted rings, each approximately half a
beam width wide, and at first all six parameters describing each ring
were left free: the center coordinates ($x_0$,$y_0$), systemic velocity
$V_{\rm sys}$,
inclination $i$, position angle of the kinematic major-axis
$\phi$ and rotation velocity $V_{\rm rot}$. After a few trials with
the velocity field of the NL~datacube, the systemic
velocity was fixed to $V_{\rm sys}=784$~\kms{},
and the center coordinates were fixed at
$(\alpha=6^{\rm h}05^{\rm m}43\fs4, \delta=-33\degr04\arcmin31\arcsec$;
J2000).

As the inclination is only loosely constrained by the observations,
we fixed it at the HyperLEDA value of $i=70\fdg5$. This value for the
inclination
depends on the assumed thickness of \eso{}, but is consistent with the
values returned by \rotcur{} in our first trials, and
is likely a good approximation of the true inclination of \eso{}.
For an oblate galaxy with semi-major axis $a$ and semi-minor axis $c$, the
relation between
the intrinsic thickness $q_0 \equiv c/a$ and the projected axis ratio $q$
is given by
\begin{equation}
\cos^2 i = \frac{q^2-q_0^2}{1-q_0^2}
\end{equation}
(e.g. \citealt{bgpv83}).
For an intrinsic thickness $q_0=0.2$ (\citealt{holmberg1946}),
an inclination $i=70\fdg5$ implies a projected
axis ratio $q=0.38$, consistent with our observations at large radii
(e.g.\ Figures~\ref{fig:opt_epa} and \ref{fig:hi_mom0}). In any case,
the rotation velocities can easily be scaled for other inclination
values.

\begin{figure}[tb]
 \includegraphics[width=0.5\textwidth]{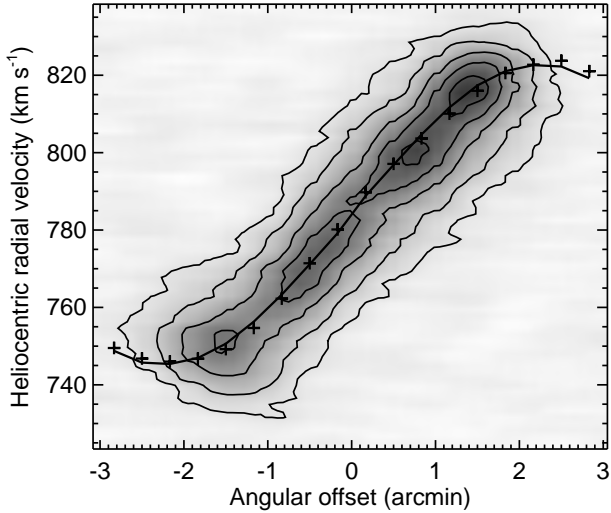}
\caption{The rotation curve of \eso{} as derived from the
 tilted-ring model described in the text, overlaid on a grayscale of
 the major-axis position-velocity diagram
 (datacube~NL; beam size $58\arcsec \times 38\arcsec$).
 The data points at radii larger than $2\farcm3$ are obtained from
 datacube~NH (beam size $132\arcsec \times 70\arcsec$), which has
 a higher sensitivity.
 The solid curve shows the fit to the whole
 velocity field, while the crosses show the fit
 to the approaching side ({\em left}) and receding side ({\em right})
 of the galaxy only.
 Contour levels are overlaid at 7.5, 20, 40, 60 and 80~mJy beam$^{-1}$.
 Rotational velocities are {\it not} corrected for the uncertain
 inclination. For an inclination $i=70\fdg5$ (HyperLEDA),
 the rotation curve must be scaled up by $6\%$.
 }
\label{fig:vrot}
\end{figure}

With the values of $(x_0,y_0)$, $V_{\rm sys}$ and $i$ are fixed,
we fit for the kinematic position angle $\phi$ and
rotation velocity $V_{\rm rot}$ as a function of semi-major axis position
$a$ using datacube NL.
The fitted position angle
is consistent with $\phi = 60\degr \pm 5\degr$ over the whole semi-major
axis range.
The resulting rotation curve is shown
in Figure~\ref{fig:vrot}, overlaid on a position-velocity diagram taken
along the major-axis. As the \hi{} column density decreases rapidly at large
radii ($a \ga 2\farcm3$), we obtain
our data points at these radii from the high-sensitivity datacube~NH.
The rotation curve is typical of late-type systems, rising in an almost
solid-body manner in the central parts. The maximum rotational
velocity of $V_{\rm rot}\sin i \approx 40$~\kms{} is reached
at $a\approx 2\arcmin$ (the optical extent), corresponding to a
projected distance of $6.3\pm 0.8 $~kpc, and the rotation curve
flattens out at larger radii.
The data hint at a decline in the rotational velocity at radii
$a \ga 2\farcm7$, though deeper \hi{} observations are necessary
to confirm this.
One should however remember that \rotcur{}
assumes axisymmetry (like {\tt Ellipse} for the surface
photometry; see \S~\ref{sec:surf_phot}), which may not be the case in
\eso{}. Our
results may thus be systematically biased.

%
%
\section{Distance Determination and Total Fluxes\label{sec:dist_fluxes}}

No precise determination of the distance to \eso{} is
available in the literature (e.g.\ through Cepheids or the tip of the
red-giant branch), but it can be estimated in several ways.
Adopting $H_0=70$~\kms{} and a Local Group infall toward the Virgo cluster
of $208$~\kms{}, HyperLEDA lists a distance of $8.02$~Mpc.
For our analysis we adopt the slightly more refined linear Virgocentric flow
model of
\citet{s80}. We derive a distance to \eso{} of $D=10.8\pm1.4$~Mpc
(random errors only), adopting a Virgo distance of $D_{\rm
Virgo}=16.1\pm1.2$~Mpc
(random errors only; \citealt{ketal00}), an observed Virgo redshift
of $V_{\rm Virgo}=1079$~\kms{} (\citealt{eetal98}), a Local Group
Virgocentric infall velocity of $257\pm32$~\kms{},
(\citealt{ahmst82}) and a Virgocentric density contrast $\gamma=2$
(representative of Abell clusters; \citealt{p76}).
An angular separation of
$1\arcmin$ then corresponds to a projected distance $3.1\pm0.4$~kpc.

At this adopted distance, the total \hi{} flux $F_{\rm
HI}=23.1\pm1.2$~\jykms{}
(see \S~\ref{sec:hi_datared})
corresponds to a total \hi{} mass $M_{\rm HI}=(6.4\pm1.7)\times10^8$~$\msun$.
> From the corrected total apparent
magnitudes listed in \S~\ref{sec:surf_phot} (corrected for Galactic
extinction but not internal absorption), we derive the following
corrected absolute magnitudes: $M_B=-16.6\pm0.3$, $M_V=-16.8\pm0.3$
and $M_I=-17.9\pm0.3$~mag. This is more than a magnitude fainter than the
absolute magnitude of the \lmc{} listed in HyperLEDA
($M_B=-17.9$~mag). Combining the optical and \hi{} measurements,
we obtain a (distance independent) total \hi{} mass-to-blue-luminosity ratio
$M_\hi/L_B=(0.96\pm0.14)$~$\msun$/$\lsunblue$,
roughly five times larger than that of the \lmc{}
based on HyperLEDA's entries
($M_\hi/L_B=0.18$~$\msun$/$\lsunblue$). This is consistent with
\eso{} being of later type than the \lmc{}, although the total
\hi{} flux of the \lmc{} is highly dependent on the
area considered, given the presence of the Magellanic bridge
(e.g.\ \citealt{puetal98,bruns2005}).
%
%
\section{Discussion\label{sec:discussion}}
%
%
\subsection{Structure of \eso{} and a Comparison to the
\lmc{}\label{sec:structure}}

Figures~\ref{fig:dss}--\ref{fig:opt_images} reveal an irregular but
characteristic optical morphology for Magellanic-type dwarf
galaxies. Indeed, the main body of \eso{} exhibits a
bar-like morphology, from which protrudes a one-arm spiral circling
the galaxy by at least $90\degr$, possibly more. This structure is
sharpest in the bluer bands, suggesting that it is associated with
relatively young stars and possibly ongoing star
formation. Narrow-band \halpha{} imaging would be particularly useful
to confirm this latter point. Despite the irregular morphology,
however, the azimuthally-averaged optical profiles are approximately
exponential up to $a\approx135\arcsec$, after which they fall off more
rapidly (see Figure~\ref{fig:opt_sbp}).

In our total \hi{} map of datacube UL, the \hi{} distribution has a
radial extent of at least $2\farcm8$ (corresponding to a column
density of $0.4\times10^{20}$~cm$^{-2}$; see Figure~\ref{fig:hi_mom0}),
thus extending to at least $2.2~\rtf$ or $1.2~\rhol$. This
is fairly typical of dwarf irregular galaxies (e.g. \citealt{h97}).
The \hi{} is also asymmetric, the column density
falling off more steeply to the South-East than the North-West, and the
absolute
peak being slightly offset to the North-East. Both
effects are common in late-type dwarf galaxies (see, e.g.,
\citealt{sahs02}).

The \hi{} distribution is also slightly offset compared to the
optical in all our maps, being shifted toward the North-East, and the
\hi{} kinematic center does not correspond to the optical center as
defined by the surface photometry. The latter displacement is about
$23\arcsec$, corresponding to a projected distance of $1.2 \pm 0.2$~kpc.
This effect is present in a number of Magellanic-type
galaxies (e.g.\ \citealt{vf72}). In particular, there are strong
indications in the \lmc{} that the kinematic center derived from
the \hi{} (e.g.\ \citealt{ketal98}) is very different from that
of the main disk, as traced by the distribution of red giant branch
(RGB) and asymptotic giant branch (AGB) stars (\citealt{m01}) or the
kinematics of intermediate-age carbon stars (\citealt{mahs02};
\citealt{olsen2007}), while
the morphological centers derived from those latter tracers are in
rough agreement with each other (\citealt{wn01}). Evidence is also
gathering that the \lmc{} is intrinsically elongated
(\citealt{mc01}; \citealt{marel2006}), presumably due to the tidal force
from the Milky Way
(\citealt{m01}; \citealt{mastropietro2005}).
For a distant galaxy, this would translate in
different kinematic and photometric major axes.
Given the uncertainties in the position angle, however, no significant
difference is seen between the photometric and kinematic position
angle profiles of \eso{} (Figures~\ref{fig:opt_epa} and
\ref{fig:hi_mom1}).

The rotation curve of \eso{} rises smoothly and reaches
a value of about $V_{\rm rot} \sin i = 40$~\kms{} in the outer parts
(Figure~\ref{fig:vrot}). In comparison, the rotation of the \lmc{}
shows a sudden decline after its peak velocity $V_{\rm rot}
\approx63$~\kms{}
at $a=2.4$~kpc  (\citealt{ketal98}).
An inclination of about $40\degr$ is expected if
\eso{} is to have the same rotation as the
\lmc{}. This however appears unlikely given the relatively high apparent
ellipticity observed (Figures~\ref{fig:opt_epa} and
\ref{fig:hi_mom0}). Even the lowest optical ellipticity measured
($\epsilon_{\rm min}=0.4$; see Figure~\ref{fig:opt_epa}) is inconsistent
with such a low inclination, for any intrinsic disk thickness (e.g.\
\citealt{bgpv83}). As a disk with an intrinsically high ellipticity is
unlikely, it is almost certain that \eso{}
is less massive than the \lmc{} over the range of radii probed.
Interestingly however, the total \hi{} mass of
\eso{} is about a third higher than that of the
\lmc{} ($\approx (4.8\pm0.2)\times10^8$~$\msun$;
\citealt{skchk03}), resulting in a significantly larger
\hi{} mass-to-blue-luminosity ratio
$M_\hi/L_B=(0.96\pm0.14)$~$\msun$/$\lsunblue$
for \eso{} (see \S~\ref{sec:dist_fluxes}).

Despite a number of similarities between the optical and \hi{}
content of \eso{} and the \lmc{}, one large
and surprising difference exists. The medium- and large-scale
\hi{} structure of the \lmc{} is very regular, bearing
little resemblance to the optical structure and lacking any obvious trace of
the
main bar (e.g.\ \citealt{puetal98,ketal03}). As stated above, however,
the \hi{} distribution of \eso{} is highly
asymmetric and roughly follows its optical morphology. In particular,
and as best shown in the \hi{} map (Figure~\ref{fig:hi_mom0}), the \hi{} is
concentrated along the main bar and appears to follow the one-arm
spiral to the North-West, in addition to a third intermediate component
(also
with an optical counterpart, at least partially). We have no
explanation for these differences in the \hi{} morphology; further
observations
in ATCA $1.5$~km configurations are needed to characterize the
\hi{} morphology of \eso{} better. In this
respect, it would be interesting to see if the optical morphology of
\eso{} remains the same at longer wavelengths or if,
like the \lmc{}, the preponderance of the bar decreases and the
smoothness of the disk increases in the near-infrared (e.g.\
\citealt{m01}, \citealt{meixner2006}). Also of interest is
whether the \hi{} differences arise
from internal processes (e.g.\ supernovae feedback) or from external
factors (e.g.\ tidal interactions). The former appears to dominate the
\hi{} structure of the \lmc{} on small scales
(\citealt{ketal98}; \citealt{ketal03}) and the latter on large scales
(\citealt{puetal98}; \citealt{skchk03}). However, this could be different
for \eso{}.

Interestingly, the velocity dispersion maps of \eso{}
and the \lmc{} are also somewhat different. In
\eso{}, the highest \hi{} velocity
dispersions are observed along the bar and the third component. In the
\lmc{}, the \hi{} velocity dispersion is higher
along the Eastern edge of the \object{30~Doradus} region and near the dense
molecular clouds extending to the South (\citealt{ketal03}). Both regions
are turbulent due to the combined effects of shocks from ram pressure
on the Galactic halo and the very active star formation near
\object{30~Doradus}. Star formation is probably also at the origin of the high
\hi{} velocity dispersions observed near the North-West end of
the \lmc{} bar and Constellation III (e.g. \ \citealt{efremov1998}).

%
%
\subsection{Comparison to Other Dwarf Irregulars\label{sec:dirrs}}

%
The optical and \hi{} distributions of \eso{} are typical of those of a
Magellanic dwarf galaxy: a prominent bar and a pronounced single spiral arm
in the optical,  an \hi{} distribution extending to about 1.2~Holmberg radii
and a kinematic center offset from the photometric center. Furthermore, our
\hi{} data suggest the presence of numerous shells and holes (see
\S~\ref{sec:shells}).

The \hi{} kinematic center of \eso{} is offset by a projected distance of
$1.2\pm 0.2$~kpc from the optical center (as derived from surface
photometry). This offset is known to be present in other Magellanic-type
systems (e.g.\ \citealt{vf72}) such as \object{NGC~925}, which has an offset of
$\approx 1$~kpc (\citealt{pisano1998}), and the \lmc{} (see
\S~\ref{sec:structure}).
Additional features in the \hi{} distribution have been detected in
Magellanic dwarfs, such as loops surrounding the galaxy (\object{NGC~4618};
\citealt{bush2004}), external spurs or blobs (e.g. \object{NGC~5169};
\citealt{muhle2005}; \object{UGCA~98}; \citealt{stil2005}) and S-shaped distortions
in the \hi{} velocity field (e.g. \object{NGC~4449}; \citealt{hunter1998}; \object{DDO~43};
\citealt{simpson2005}). Such features, often associated with a recent or
ongoing interaction, are not apparent in the \hi{} distribution of \eso{}.
The origin of the third component identified in \hi{} between the optical
bar and one-arm spiral also remains unclear.
Shell- and hole-like structures in the \hi{} distribution, as observed in
\eso{} (see \S~\ref{sec:shells}), are observed in practically all dwarfs.
These structures are commonly thought to be the result of stellar winds and
supernova explosions (e.g. \citealt{weaver1977,mccray1987,pwbr92,ketal99};
but see also \citealt{rswr99,bc02,dib2005}).

We find no significant difference between the photometric and kinematic
position angle of \eso{}. \cite{hunter2000} find a similar result in their
\hi{} study of \object{UGC~199}. Many other Magellanic dwarfs, however, show a
significant disagreement between the morphological and kinematic axes, such
as \object{NGC~1156} (\citealt{hetal2002}) and \object{DDO~26} (\citealt{hw02}). \object{DDO~43} even
has a kinematic axis nearly perpendicular to its morphological axis
(\citealt{simpson2005}). \cite{hetal2002} suggest that such apparent
inconsistencies may be explained by the presence of an inclined bar.

The asymmetry we observe in the global \hi{} profile of \eso{}
is a common characteristic of Magellanic-type galaxies,
often attributed to a recent or ongoing interaction.
\cite{bush2004}, for example, analyse the \hi{} morphology
of the interacting Magellanic spiral galaxies
\object{NGC~4618} and \object{NGC~4625}, particularly in relation to the interaction between
them.
Their \hi{} observations show a loop-like structure around \object{NGC~4618} (the
most
massive of the two), indicating that the outer gas of this galaxy
is strongly perturbed by the recent interaction.
\object{NGC~4625} and the inner part of \object{NGC~4618}, however, appear unaffected by the
interaction.
They find an asymmetry ratio $A=1.0$ for the \hi{} profile of \object{NGC~4618}
and $A=1.29$ for \object{NGC~4625}, where $A$ is defined as the ratio between
the areas under the \hi{} profile at velocities smaller and greater
than the systemic velocity (\citealt{haynes1998}).
Using the same algorithm,
we find an asymmetry ratio $A=1.1$ for \eso{}, which is bracketed by the
values for \object{NGC~4618} and \object{NGC~4625}. However, a similar amount of asymmetry
is detected in non-interacting Magellanic spirals (\citealt{wilcots2004});
an asymmetry measurement alone cannot confirm or reject the hypothesis of
a recent interaction.

\eso{} exhibits solid-body rotation at small radii, reaches its maximum
rotational velocity of $V_{\rm rot} \sin i \approx 40$~\kms{} at a projected
distance of $\approx 6.4$~kpc from its center, and flattens out at large
radii. This behaviour is similar to that of other Magellanic dwarfs, e.g.
\object{IC~10} ($\approx 35$~\kms{} at 1~kpc; \citealt{wilcots1998}), \object{DDO~43}
($\approx 30$~\kms{} at 2~kpc; \citealt{simpson2005}) and \object{NGC~4618} ($\approx
50$~\kms{} at 5~kpc; \citealt{bush2004}), though variations in these
parameters clearly exist.
The \hi{} velocity dispersion of \eso{} is $8$--$16$~\kms{}
over most of the galaxy (Figure~\ref{fig:hi_mom2}), which is very similar to
that of other dwarfs.
The Magellanic dwarfs mentioned above exhibit a somewhat higher
velocity dispersion at several locations (including the bar), indicating
recent or
ongoing star formation. For \eso{}, the largest velocity
dispersion is reached over the optical bar and the third component,
which also likely indicates recent star formation in these regions.

\cite{doyle2005} present a search for optical counterparts of HIPASS
sources. Out of the 3618 optical counterparts found, 151~are galaxies with
morphological type IB(s)m, including \eso{}. From their (total) \hi{} fluxes
and $B$-band magnitudes we calculated the \hi{} mass-to-blue-luminosity
ratio $M_\hi/L_B$ for each galaxy. Magellanic dwarf irregular galaxies have
a median value $\langle M_\hi/L_B \rangle =0.97$~$\msun / \lsunblue$, with a
standard deviation of $0.74$~$\msun / \lsunblue$. Thus, \eso{} (with
$M_\hi/L_B=(0.96\pm0.14)$~$\msun / \lsunblue$) is a fairly typical Magellanic
dwarf galaxy in this respect as well.

%
\subsection{Group Membership and Origin of the
Asymmetries\label{sec:group_m=1}}

Tidal interactions in disk galaxies are known to lead to strongly
asymmetric spiral features (e.g.\ \citealt{bsv86,ok90}). The structure of
\object{M51}-like
systems, for example, is often explained by tidal interactions with
a close neighbour (e.g.\ \citealt{hb90,hkbb93,sl00}). While this
mechanism need not be unique, it is interesting to question whether the
large-scale morphology of \eso{} is consistent with a
(trans)formation through tidal interactions, or whether another mechanism
needs to be invoked. Certainly, the break in \eso{}'s light
profile seen at $a\approx135\arcsec$ is consistent with an
interaction, since tidal features typically have sharp boundaries.

\citet{t88} lists \eso{} as being part of a small group
of $3$ galaxies in the Dorado Cloud, which also includes \object{NGC~2090}
and \object{NGC~2188}. Table~\ref{tab:group} lists all galaxies
with a known redshift within a radius of $5\degr$ of \eso{},
and with a relative radial velocity less than $500$~\kms{},
as found in NED. Only one additional galaxy is
found, \object{AM~0605-341}, although a number of galaxies without
redshift are also present in the $5\degr$ region.
There are thus a number of candidates for
interaction with \eso{}.

\begin{table}[tb]
 \caption{Basic properties of galaxies near \eso{}. This table lists
 all galaxies with a projected distance differing by less than $5\degr$ and
 a redshift differing by less than 500~\kms{} from \eso{}. Only galaxies
with
 a known redshift are included.
 From left to right, the
 columns list the galaxy name, spectral type, total apparent $B$ magnitude
 and the angular separation, projected separation
 (assuming a distance of 10.8~Mpc), and relative radial velocity from
\eso{}. }
 \label{tab:group}
 \begin{tabular}{llcrcc}
   \hline
   \hline
    Name & Type & $B_{\rm T}$ & $\Delta\theta$ & $\Delta R$ & $\Delta V_{\rm
sys}$ \\
   &      &       (mag) &                &        (kpc) &        (\kms{}) \\
   \hline
   \eso{}                &  IB(s)m & $13.8$ &          --- &   --- &    ---
\\
   \object{AM~0605-341}  &    SBdm & $14.1$ &  $70\farcm2$ & $221$ &  $-19$
\\
   \object{NGC~2188}     &  SB(s)m & $12.1$ &  $82\farcm5$ & $259$ &  $-37$
\\
   \object{NGC~2090}     & SA(rs)b & $12.0$ & $244\farcm0$ & $768$ & $+137$
\\
   \hline
 \end{tabular}\\
\end{table}

\object{NGC~2090} is the least certain and most distant
member of the group containing \eso{} (among the galaxies
with a measured redshift). Its distance was measured using the Hubble
Space Telescope (HST) and the Cepheid period-luminosity relation,
yielding a distance of $12.3\pm0.9\pm0.9$~Mpc (random and
systematic errors, respectively; \citealt{peetal98}). Although slightly
higher, this
is consistent within the errors with our adopted distance for
\eso{}, a comforting fact given that our estimate for \eso{} was
itself higher than other published values (see
\S~\ref{sec:dist_fluxes}).

The edge-on galaxy \object{NGC~2188} is rather peculiar, both the
\hi{} and \halpha{} emission bending away from the disk,
leading to a crescent shape not unlike that expected from ram pressure
stripping (\citealt{ddd96}). The \hi{} distribution is
also highly asymmetric with respect to the equatorial plane. The
velocity field shows many peculiarities, with apparent rotation about
two axes. Although not bullet-proof evidence for
interaction, these properties hint at least to a disturbed past. Given
the large separation between \eso{} and \object{NGC~2188}, however, any
interaction with
\eso{} would have occurred a very long time ago
($>1.25$~Gyr for a relative velocity of $200$~\kms{}). The
evidence for a recent interaction involving \eso{} thus
remains marginal.
%
%
\subsection{\hi{} Shells and Holes\label{sec:shells}}

Although the current observations are best suited to study the medium
and large-scale structure of the \hi{}, studies of the
small-scale structure in the interstellar medium (ISM) are also of
interest. In particular, the radiation and mechanical energy produced
by stellar winds and supernova explosions are generally thought to
give rise to a very dynamic multi-phase ISM, through the interacting
and continually evolving cavities created. For recent work on the latter
topic,
see, e.g., \cite{kbstn99} and \cite{ab01}. Such shells and holes have indeed
been
observed in nearby large spiral galaxies such as \object{M31} (e.g.\
\citealt{bb86}) and M33 (\citealt{dh90}). Dwarf galaxies
are however better targets for studying these phenomena.
The low gravitational potential well of dwarf
galaxies, the lack of shear due to solid-body rotation and the absence
of density waves all facilitate the expansion and survival of
the shells created. Some of the best examples include the \lmc{}
(SB(s)m; \citealt{ketal98}), \object{IC~2574} (SAB(s)m;
\citealt{wb99}) and \object{Holmberg~II} (Im; \citealt{pwbr92}; but
see also \citealt{rswr99,bc02}).

The channel maps of datacube~UL
(Figure~\ref{fig:chan_maps}) and the corresponding moment maps
(Figures~\ref{fig:hi_mom0}--\ref{fig:hi_mom2}) of \eso{}
reveal much small-scale structure, suggesting that expanding
shells or holes may be present in \eso{}. Most obvious are the two large
\hi{}
gaps, the first one between the main bar and the so-called
third component, centered at ($\alpha=06^{\rm h}05^{\rm m}44\fs5$,
$\delta=-33\degr04\arcmin10\arcsec$; J2000), and the second one
between the third component and the one-arm spiral, itself perhaps
composed of two smaller holes centered respectively at
($\alpha=06^{\rm h}05^{\rm m}38\fs7$,
$\delta=-33\degr04\arcmin50\arcsec$) and ($\alpha=06^{\rm h}05^{\rm
m}39\fs9$, $\delta=-33\degr05\arcmin20\arcsec$). These holes raise the
interesting possibility that the third component and the one-arm
spiral may not be caused by large-scale dynamical processes, but may
instead simply represent gas concentrations at the edges of large
cavities, emptied by the supernova explosions and stellar winds
associated with past bursts of star formation. Their optical
counterparts would then trace the location of secondary (i.e.\
triggered and more recent) star formation.

Despite its relatively large distance, \eso{} is thus a
prime target for studies of star formation feedback and
self-propagating star formation. However, due to the short
integrations with the 1.5D array and the shallowness of the
high-resolution maps, it is hard to convincingly argue that any single
structure is surely due to a (centrally-located) localized energetic
phenomenon. Better quality maps and longer observations with $1.5$~km
or longer arrays are necessary for a proper study of the small-scale
structure of the ISM.
As mentioned before, narrow-band \halpha{} imaging
would be useful to see how current and recent sites of star formation
relate to the \hi{} morphology and kinematics.
%
%
\section{Summary and Conclusions\label{sec:summary}}

We presented the analysis of optical imaging observations and \hi{} radio
synthesis observations of the
dwarf irregular galaxy \eso{}. The optical $BVI$ imaging data reveal a
morphology characteristic
of Magellanic-type spirals
and irregulars, with a large dominant bar and a one-sided spiral or
tidal arm, although the absolute magnitude of \eso{} (for a distance
$D=10.8\pm1.4$~Mpc) is more than a magnitude
fainter than that of the \lmc{}. While poorly-defined, the
azimuthally-averaged
surface brightness profiles show an exponential disk with a possible
break at large radii and colors typical of late-type disk galaxies.

The radio synthesis observations reveal
an \hi{} disk extending well outside of the optical
extent. The total \hi{} mass is $M_\hi=(6.4\pm 1.7)\times10^8$~$\msun$,
yielding
a (distance independent) \hi{} mass-to-blue-luminosity ratio
$M_\hi/L_B=(0.96\pm0.14)$~$\msun$/$\lsunblue$,
significantly more \hi{}-rich than the \lmc{}. The latter
value is at the high end of the distribution for late-type spirals
(Sc--Sd), but is typical of late-type dwarfs
(\citealt{rh94,sahs02,doyle2005}). The large-scale \hi{}
distribution is also asymmetric and roughly follows the optical light
distribution,
although slightly offset from it. The \hi{} distribution
thus appears consistent with that expected from tidal interactions,
but most evidence for a past encounter is circumstantial. Our highest
spatial resolution maps show that the highest column densities and velocity
dispersions
are reached over the central bar,
the one-arm spiral, and over a third component which also has an
optical counterpart (at least partially). Despite a similar optical
morphology, this is opposite to what is observed in the \lmc{},
where the \hi{} distribution bears little resemblance to that of the
stars. The two galaxies may thus have formed differently, or may
simply be in a different evolutionary stage. The rotation curve of
\eso{} is solid-body over the optical extent but
flattens out at large radii, reaching $V_{\rm rot}\sin i \approx 40$~\kms{}.
The inclination remains poorly constrained by observations, but is
consistent
with the HyperLEDA value $i \approx 70\fdg5$.
Our high-resolution observations also hint at a complex \hi{}
structure, reminiscent of that expected from stellar winds and
supernova explosions. Deeper and higher-resolution
optical and \hi{} observations, as well as \halpha{} observations,
are however necessary to properly characterize the small-scale \hi{}
morphology
and kinematics, and to establish any relationship to star formation.

The observations presented in this paper provide important information on
the
large-scale optical and \hi{} structure of
\eso{}, and provide a solid foundation for further
investigation of its small-scale structure, particularly in
\hi{} with longer baseline observations. The ultimate
goal is that studies of individual nearby Magellanic galaxies will
yield information on their formation and evolution as a class, but
will also strengthen our understanding of the \lmc{} itself.

%
%
\begin{acknowledgements}
It is a pleasure to thank Claude Carignan and Ken Freeman 
for their support in the initial phase of the work reported here.
TK was supported by NWO under project number 614.041.006 and by 
PPARC under grant number PP/D002036/1. 
MB acknowledges support from NASA through Hubble Fellowship grant
HST-HF-01136.01 awarded by Space Telescope Science Institute, which is
operated by the Association of Universities for Research in Astronomy,
Inc., for NASA, under contract NAS~5-26555, during much of this
work. MB also acknowledges support from the Astrophysical Research
Center for the Structure and Evolution of the Cosmos (ARCSEC) at
Sejong University while this manuscript was prepared. 
SK was supported in part by Korea Science \& Engineering Foundation (KOSEF)
under a cooperative agreement with the Astrophysical Research Center of the
Structure and Evolution of the Cosmos (ARCSEC).
The NASA/IPAC
Extragalactic Database (NED) is operated by the Jet Propulsion
Laboratory, California Institute of Technology, under contract with
the National Aeronautics and Space Administration. 
This research made
use of HyperLEDA: http://leda.univ-lyon1.fr\,. The Digitized Sky
Surveys were produced at the Space Telescope Science Institute under
U.S. Government grant NAG W-2166. The images of these surveys are
based on photographic data obtained using the Oschin Schmidt Telescope
on Palomar Mountain and the UK Schmidt Telescope.
\end{acknowledgements}
%
%

\end{document}